\documentclass[3p,preprint]{elsarticle}

\usepackage{amssymb}

\usepackage[figuresright]{rotating}
\usepackage{subcaption}

\newtheorem{theorem}{Theorem}
\newtheorem{lemma}[theorem]{Lemma}
\newtheorem{proposition}[theorem]{Proposition}
\newtheorem{corollary}[theorem]{Corollary}
\newtheorem{definition}[theorem]{Definition}
\newtheorem{claim}{Claim}
\newtheorem{example}{Example}
\newtheorem{remark}{Remark}

\newproof{proof}{Proof}
\newproof{proofsketch}{Proof Sketch}

\usepackage[utf8]{inputenc}
\usepackage{complexity}
\usepackage[algosection]{algorithm2e}
\usepackage{endnotes}
\usepackage{listings}
\usepackage{mathtools}
\usepackage{verbatim}
\usepackage{mathrsfs}
\usepackage{longtable}
\usepackage{enumitem}
\usepackage{etoolbox}
\usepackage{float}
\usepackage{color}
\usepackage{hyperref}

\usepackage{complexity}
\usepackage{comment}
\usepackage{longtable}

\usepackage{mathtools}

\DeclareMathOperator{\amb}{{\rm amb}}

\newcommand*{\DEBUG}{}%

\ifdefined\DEBUG
\newcommand{\fixme}[1]{{\textcolor{red}{\bf{\textsf{FIXME: #1}}}}}
\newcommand{\bug}[1]{{\textcolor{blue}{\bf{\textsf{BUG: #1}}}}}
\newcommand{\idea}[1]{{\textcolor{blue}{\bf{\textsf{IDEA: #1}}}}}

\newcommand{\TODO}[1]{{\textcolor{red}{\bf{\textsf{ 
TODO: #1
}}}}}
\else
\newcommand{\fixme}[1]{}
\newcommand{\bug}[1]{}
\newcommand{\TODO}[1]{}
\newcommand{\idea}[1]{}
\fi

\newcommand{\derivestep}[1][]{\mathrel{\mathop{\Rightarrow}_{#1}}}
\newcommand{\uuline}[1]{\underline{\underline{#1}}}

\newclass{\COMSLIP}{COM\mbox{-}SLIP}
\newclass{\COMSLIPCUP}{COM\mbox{-}SLIP^{\cup}}

\newclass{\DCM}{DCM}
\newclass{\eDCM}{eDCM}
\newclass{\eNPDA}{eNPDA}
\newclass{\DPDA}{DPDA}
\newclass{\RDPDA}{RDPDA}
\newclass{\PDA}{PDA}
\newclass{\DCMNE}{DCM_{NE}}
\newclass{\TwoDCM}{2DCM}
\newclass{\NCM}{NCM}
\newclass{\eNCM}{eNCM}
\newclass{\eNQA}{eNQA}
\newclass{\eNSA}{eNSA}
\newclass{\eNPCM}{eNPCM}
\newclass{\eNQCM}{eNQCM}
\newclass{\eNSCM}{eNSCM}
\newclass{\DPCM}{DPCM}
\newclass{\NPCM}{NPCM}
\newclass{\NQCM}{NQCM}
\newclass{\NSCM}{NSCM}
\newclass{\NPDA}{NPDA}
\newclass{\TRE}{TRE}
\newclass{\NFA}{NFA}
\newclass{\DFA}{DFA}
\newclass{\NCA}{NCA}
\newclass{\DCA}{DCA}
\newclass{\DTM}{DTM}
\newclass{\NTM}{NTM}
\newclass{\DLOG}{DLOG}
\newclass{\CFG}{CFG}
\newclass{\ETOL}{ET0L}
\newclass{\EDOL}{ED0L}
\newclass{\EOL}{E0L}
\newclass{\EDTOL}{EDT0L}
\newclass{\CFP}{CFP}
\newclass{\ORDER}{O}
\newclass{\MATRIX}{M}
\newclass{\BD}{BD}
\newclass{\LB}{LB}
\newclass{\ALL}{ALL}
\newclass{\decLBD}{decLBD}
\newclass{\StLB}{StLB}
\newclass{\SBD}{SBD}
\newclass{\TCA}{TCA}

\newtheorem{problem}{Problem}

\newsavebox{\spacebox}
\begin{lrbox}{\spacebox}
\verb*! !
\end{lrbox}
\newcommand\natzero{\mathbb{N}_0}
\newcommand\Nc{\widehat{\mathbb N}}

\newcommand{\LL}{{\cal L}}

\newcommand\bd{^{{\rm bd}}}
\newcommand\fin{_{{\rm fin}}}

\begin{document}

\begin{frontmatter}




\title{Relationships Between Bounded Languages, Counter Machines, Finite-Index Grammars, Ambiguity, and Commutative Regularity\tnoteref{t1}}

\tnotetext[t1]{\textcopyright 2022. This manuscript version is made available under the CC-BY-NC-ND 4.0 license \url{http://creativecommons.org/licenses/by-nc-nd/4.0/} The manuscript is published in A. Carpi, F. D'Alessandro, O.H. Ibarra, I. McQuillan. Relationships Between Bounded Languages, Counter Machines, Finite-Index Grammars, Ambiguity, and Commutative Regularity. {\it Theoretical Computer Science} 862, 97--118 (2021).}

\tnotetext[t1]{This paper includes selected results from (O.H.\ Ibarra, I.\ McQuillan, On Bounded Semilinear Languages, Counter Machines, and Finite-Index ET0L, Proceedings of the Conference on Implementation and
Application of Automata (CIAA 2016), Lecture Notes in Computer Science 9705). It also contains new results not in the proceedings, such as Section \ref{sec:Finite-Index-ET0L-amb} onwards.}

\author[label1]{Arturo Carpi\corref{corr}}
\address[label1]{Dipartimento di Matematica e Informatica\\  Universit\`a degli Studi di Perugia,
Perugia, Italy}
\ead[label1]{carpi@dmi.unipg.it}
\cortext[corr]{Corresponding author}

\author[label2]{Flavio D'Alessandro\fnref{fn3}
\corref{corr}}
\address[label2]{Department of Mathematics\\ Sapienza University of Rome,
00185 Rome, Italy
 \\ and		\\
  Department of Mathematics, 
Bo\u gazi\c ci University \\
  34342 Bebek, Istanbul, Turkey}
\ead[label2]{dalessan@mat.uniroma1.it}
\fntext[fn3]{Supported, in part, by TUBITAK Project 2221, Scientific and Technological Research Council of Turkey, (Flavio D'Alessandro)}

\author[label3]{Oscar H. Ibarra\fnref{fn1}
\corref{corr}}
\address[label3]{Department of Computer Science\\ University of California, Santa Barbara, CA 93106, USA}
\ead[label3]{ibarra@cs.ucsb.edu}
\fntext[fn1]{Supported, in part, by
NSF Grant CCF-1117708 (Oscar H. Ibarra)}

\author[label4]{Ian McQuillan\fnref{fn2}
\corref{corr}}
\address[label4]{Department of Computer Science, University of Saskatchewan\\
Saskatoon, SK S7N 5C9, Canada}
\ead[label4]{mcquillan@cs.usask.ca}
\fntext[fn2]{Supported, in part, by Natural Sciences and Engineering Research Council of Canada Grant 2016-06172 (Ian McQuillan)}

\begin{abstract}
It is shown that for every language family that is a trio
containing only semilinear languages, all
bounded languages in it can be accepted by
one-way deterministic reversal-bounded multicounter
machines ($\DCM$). This implies that
for every semilinear trio (where these properties are effective), it
is possible to decide containment, equivalence, and disjointness
concerning its bounded languages. 
A condition is also provided
for when the bounded languages in a semilinear trio
coincide exactly with those accepted by $\DCM$ machines,
and it is used to show that many grammar systems of finite index --- such as finite-index matrix grammars 
($\M\fin$) and finite-index $\ETOL$ ($\ETOL\fin$) --- have identical bounded languages as $\DCM$. 

Then connections between ambiguity, counting regularity, and commutative regularity are made, as many machines and grammars
that are unambiguous can only generate/accept counting regular or commutatively regular languages. Thus, such a system
that can generate/accept a non-counting regular or non-commutatively regular language implies the existence of inherently ambiguous
languages over that system. In addition, it is shown that every language generated by an unambiguous $\M\fin$ has a rational characteristic series in commutative variables, and is counting regular. This result plus the connections are used to demonstrate that the grammar systems $\M\fin$ and $\ETOL\fin$ can generate inherently ambiguous languages (over their grammars), as do several machine models. It is also shown that all bounded languages
generated by these two grammar systems (those in any semilinear trio) can be generated unambiguously within the systems.
Finally, conditions on $\M\fin$ and $\ETOL\fin$ languages implying commutative regularity are obtained. In particular, it is shown that every finite-index $\EDOL$ language is commutatively regular. 
\end{abstract}


\end{frontmatter}

\section{Introduction}
\label{sec:intro}


The notions of bounded languages and semilinear sets and
languages are old ones in the area of formal languages 
(see e.g.\ \cite{GinsburgCFLs}),
and they have been used and applied extensively.
To recall, a language $L\subseteq \Sigma^*$ is {\em bounded} if there
exist words, $w_1, \ldots, w_k \in \Sigma^+$, such that 
$L\subseteq w_1^* \cdots w_k^*$.
The formal definition of semilinear sets
$Q \subseteq \mathbb{N}_0^k$ appears in Section \ref{sec:prelims}, and
a language is semilinear if its Parikh
image is a semilinear set (equivalently, a language is semilinear
if and only if it has the same Parikh image as some regular language \cite{harrison1978}). Many well-studied language families, such as the context-free
languages, only contain languages that are semilinear \cite{Parikh}.

In the formal language theory literature, when creating a new machine model
or grammar system, it is common to investigate decision problems, such as the decidability of
the membership problem (is $w \in L(M)$?), the emptiness problem (is $L(M) = \emptyset$?),
the equivalence problem (is $L(M_1) = L(M_2)$?), the containment problem
(is $L(M_1)\subseteq L(M_2)$?), and the disjointness problem
(is $L(M_1) \cap L(M_2) = \emptyset$?). When some of these
problems are undecidable, it is common to then study these problems for the special case over just the bounded languages in these families. However, there is not any sort of general strategies for
studying these decision properties on bounded languages from these different
models. Here, such a strategy is obtained that even allows a comparison across
two families at once.

There are various ways of combining the notions of boundedness and semilinearity.
In this
paper, four different ones are considered and compared.
In particular, a language $L \subseteq \Sigma^*$ 
has been called
{\em bounded semilinear} if there exists a semilinear set $Q \subseteq \natzero^k$ and words 
$w_1, \ldots, w_k$ such that such that
 $L =  \{ w  \mid w =  w_1^{i_1} \cdots w_k^{i_k}, (i_1, \ldots, i_k) \in Q\}$
\cite{IbarraSeki}.
In this paper, we refer to these bounded semilinear languages as {\em
bounded Ginsburg semilinear} to disambiguate with other types.
Here, we provide three other definitions combining the notions of boundedness and semilinearity and compare them.
It is already known that the bounded Ginsburg semilinear languages are exactly the bounded languages that can be accepted by a one-way nondeterministic reversal-bounded
multicounter machine ($\NCM$) \cite{Ibarra1978}. 
Furthermore, it is known that every
$\NCM$ machine accepting a bounded language can be converted
to a deterministic machine ($\DCM$) accepting the same language
\cite{IbarraSeki}.

In this paper, it is demonstrated that for every
semilinear
trio $\LL$ (a family where all languages are semilinear, and 
is closed under $\lambda$-free morphism, inverse morphism, and
intersection with regular languages), every bounded language in $\LL$ is
bounded Ginsburg semilinear. Hence, all bounded
languages in any semilinear trio are in $\DCM$.
This immediately provides all of the positive decidability results for
all bounded languages in any semilinear trio (assuming
all these properties have effective constructions). Furthermore, for
the decidability problems listed
above involving two languages --- equality, containment, and disjointness --- they are all decidable
for bounded languages where both languages can be in possibly different semilinear trios. Thus, ad hoc proofs of decidability for bounded languages in semilinear trios are no longer needed since these
are decidable for $\DCM$.
Examples of such language families are the context-free languages, finite-index\footnote{The restriction of {\em finite-index} on different types of grammar systems enforces that there is an integer $k$ such that, for every word in the language, there is a derivation that uses at most $k$ nonterminals in every sentential
form.} $\ETOL$ languages (denoted by $\ETOL\fin$, a family of Lindenmayer systems \cite{RozenbergFiniteIndexETOL}), the family of languages generated by finite-index matrix grammars ($\M\fin$, \cite{PaunMatrix}, which generate the same family as $\ETOL\fin$ and several other grammar systems restricted to be finite-index \cite{RozenbergFiniteIndexGrammars}), linear indexed
languages \cite{Duske}, uncontrolled finite-index indexed languages \cite{LATA2017}, multi-push-down languages \cite{multipushdown}, 
and many others \cite{Harju2002278}.
A criterion is also developed for testing whether the bounded languages within
a semilinear trio coincide exactly with those in
$\NCM$ and $\DCM$. We apply these results to 
show that the bounded languages in 
$\ETOL\fin$ (and $\M\fin$) coincide exactly with those in $\NCM$
and $\DCM$. This is interesting given how
different the two types of systems operate. Indeed, $\NCM$
is a sequential machine model that operates with multiple independent stores, and $\ETOL\fin$
is a grammar system where rules are applied in parallel.
Hence, this work establishes general relationships between machine models and grammar systems that accept only
semilinear languages, for bounded languages.

Next, relationships between machines and grammars with respect to the important notion of ambiguity and inherent
ambiguity are investigated. While there has been extensive study regarding these notions for context-free grammars, e.g.\ \cite{Parikh},
there has been far less work done thus far on other grammar and machine models.
We are interested in the problem of determining whether various classes of grammars/machines generate/accept languages
that are inherently ambiguous over their systems.  The key concepts used for this study is commutative regularity and
counting regularity, which are defined next.
 Two words are said to be  {\em commutatively equivalent} if one is obtained from the other by rearranging the letters of the word. 
   Two languages $L_1$ and $L_2$ are said to be {\em commutatively equivalent} if there exists a 
   bijection $f\colon L_1 \to L_2$ such that every word $u\in L_1$ is commutatively equivalent to $f(u)$. In the case that $L_2$ is regular, the language $L_1$ is called {\em commutatively regular}.
  The notion of commutative regularity is a stronger notion than the following two concepts:
  1) semilinearity; 2) counting regularity. 
Indeed, the semilinearity of a language $L_1$
is equivalent to the existence of a function, not necessarily bijective,  from $L_1$ to a regular language $L_2$, preserving commutative equivalence,  and vice versa.
 For the second concept, the counting function $f_L(n)$ of a language $L$ indicates the number of words of length $n$ in $L$; the counting regular languages are those languages whose counting function are rational; that is, 
a language $L_1$ is counting regular if there exists a regular language $L_2$ and a bijection from $L_1$ to $L_2$ that preserves lengths \cite{iba,BP,iba-2-journ}. Hence, if there exists a Parikh-image preserving bijection, there must exist a length-preserving one, and so all commutatively regular languages are also counting regular.
In \cite{iba-2-journ}, it was shown that all languages accepted by unambiguous nondeterministic Turing machines with
a one-way read-only input tape and a reversal-bounded read/write worktape (there's a bound on the number of changes in direction
of the read/write head) are counting regular. Hence, if there is a machine model that can be (unambiguously)
simulated by these Turing machines, and the model accepts a non-counting regular language, then this family
contains a language that is inherently ambiguous with respect to these machines. We use this to conclude that many machine models,
such as reversal-bounded pushdown automata and reversal-bounded queue automata accept inherently ambiguous languages.

We next conduct a similar analysis for finite-index grammars. 
It is proved that languages unambiguously generated by finite-index matrix grammars
have a rational characteristic series in commutative variables. 
As a consequence, one derives that all unambiguous $\M\fin$ languages are counting regular, and there are $\M\fin$
languages that are inherently ambiguous.
The previous result  is then adapted to $\ETOL\fin$ systems. 
This is obtained by introducing the new notion of reduced $\ETOL$ system,
 that allows to formulate in an appropriate way the property of ambiguity (and that of non-ambiguity) for finite-index $\ETOL$ systems.  As a consequence of these results, we are also able to find the first inherently ambiguous language over $\ETOL\fin$ in the literature. 
 This is particularly interesting, as it was conjectured by Chomsky that there are context-free languages
 that are inherently ambiguous with respect to
a class of context-free grammars ${\cal A}$, but not inherently ambiguous with respect to a class of context-free grammars ${\cal B}$  which properly
contains ${\cal A}$ \cite{Chomsky}.  
Later, Blattner \cite{Blattner} demonstrated that there is a linear context-free language which is inherently 
ambiguous with respect to linear context-free grammars, but that is not inherently ambiguous with respect to the context-free grammars.
More generally, Blattner used grammar forms and showed that for all grammar forms generating proper subsets of the
context-free languages, there is some language that is inherently ambiguous for this sub-class of the context-free grammars, but not with respect to the context-free grammars generally. Such a form includes the $k$-linear grammars \cite{GrammarForms}, which are the finite union of products
of $k$ linear context-free languages. These can describe a strict subset of the finite-index context-free grammars (also called
derivation-bounded context-free grammars) \cite{FiniteTurn}. 
Our results show that the conjecture
is also true
for classes ${\cal A}$ and ${\cal B}$ which are not context-free grammars; for example, we
show that the
conjecture holds  for ${\cal A} = \ETOL\fin$ and ${\cal B} =  \ETOL$.

Next, in \cite{di2013-code,di2013-diof,di2013-sem}, it was shown that all bounded (Ginsburg) semilinear languages are commutatively regular. Using the aforementioned results shown in this paper, this implies that all bounded languages in any semilinear trio are commutatively regular. We further show here that all of these bounded languages can be generated by an unambiguous
$\M\fin$ and an unambiguous reduced $\ETOL\fin$; this is also true for the machine model $\NCM$ as these machines can be accepted by a $\DCM$.
Furthermore, conditions have been previously found that assure that certain context-free languages of finite index are commutatively regular \cite{CD-DLT-2018}.
Similarly here, additional
conditions are provided (in addition to the bounded language case) that assures that a language generated by a finite-index matrix grammar is
commutatively regular. In fact, it is proved that all finite-index $\EDOL$  languages (languages
generated by deterministic $\ETOL\fin$ systems with one table) are commutatively regular.

Hence, this paper makes several global connections between machine and grammar models accepting/generating semilinear
languages for bounded languages, and between ambiguity and counting regularity or commutative regularity, for both bounded and non-bounded languages.

   The paper is organized as follows. 
     In Section \ref{sec:prelims} preliminaries on the main objects studied in the paper are presented. 
In Section \ref{sec:Boun-Lang-Count-Machines}, we will discuss the results relating  bounded languages with semilinear sets. 
In Section \ref{Finite-Index-BounLan}, we will show that the bounded languages in the families of finite-index $\ETOL$ systems and of $\NCM$ are identical.
Section \ref{sec:Finite-Index-ET0L-amb}  will be dedicated to the study of the property of ambiguity for finite-index $\ETOL$ systems.
In Section \ref{sec:Char-Series-Matrix-Grammar}, we will investigate the characteristic series of finite-index matrix grammars.
In Section \ref{sec:inherent}, we study many different language families, and determine the existence of inherently ambiguous languages within them.
Section \ref{sec:Commutative-Equivalence-Problem} will be concerned with conditions under which finite-index matrix grammars and finite-index $\ETOL$ systems only contain commutatively regular languages.
Finally, Section \ref{sec:conclusions} presents the conclusions and open problems.


\section{Preliminaries}
\label{sec:prelims}

We assume a familiarity with automata and formal languages.
We refer the interested reader to \cite{DassowPaun,GinsburgCFLs,harrison1978} for introductory background. 

We will fix the notation used in this paper.
Let $\Sigma$ be a finite alphabet. Then $\Sigma^*$ (respectively $\Sigma^+$) is the set of all words (non-empty words) over $\Sigma$. 
A word $w$ is any element of $\Sigma^*$, while a language 
is any subset $L$ of $\Sigma^*$. The empty word is denoted by $\lambda$.
A language $L \subseteq \Sigma^*$ is {\em bounded} if there
exists (not necessarily distinct) words $w_1, \ldots, w_k$ such
that $L\subseteq w_1^* \cdots w_k^*$. $L$ is
{\em letter-bounded} if there exists (not necessarily distinct)
letters $a_1, \ldots, a_k$ such that $L \subseteq a_1^* \cdots a_k^*$.
If $a_1, \ldots, a_k$ are distinct, then we say $L$ is {\em distinct-letter-bounded}.
Given a language family $\LL$, the subset of $\LL$
consisting of all bounded languages in $\LL$, is denoted by $\LL\bd$.

Let $\mathbb{N}$ be the set of positive integers, and $\mathbb{N}_0$ the set of non-negative integers. Let $m \in \mathbb{N}_0$. Then,
$\pi(m)$ is $1$ if $m>0$ and $0$ otherwise.
A subset $Q$ of $\mathbb{N}_0^m$ ($m$-tuples) is a {\em linear set}
if there exist vectors $\vec{v_0}, \vec{v_1}, \ldots, \vec{v_r} \in \mathbb{N}_0^m$ such that $Q = \{\vec{v_0} + i_1 \vec{v_1} + \cdots + i_r \vec{v_r} \mid i_1, \ldots, i_r \in \mathbb{N}_0\}$. Here,
$\vec{v_0}$ is called the {\em constant}, and $\vec{v_1}, \ldots, \vec{v_r}$ are called the {\em periods}. A finite union of linear sets is called a {\em semilinear set}. 

Let $\Sigma = \{a_1, \ldots, a_n\}$ be an alphabet.
The length of a word $w \in \Sigma^*$
is denoted by $|w|$. For $a\in \Sigma$, $|w|_a$ is the number of $a$'s in $w$, and for any subset $X$ of $\Sigma$, 
$|w|_X = \sum_{a \in X} |w|_a$.
The {\em Parikh image} of $w$ is the vector 
$\psi(w) = (|w|_{a_1}, \ldots, |w|_{a_n})$, which is extended to
languages, by $\psi(L) = \{ \psi(w) \mid w \in L\}$. A language
is {\em semilinear} if its Parikh image is a semilinear set.
It is known that a language $L$ is semilinear if and only if it has
the same Parikh image as some regular language \cite{harrison1978}.
Furthermore, a language family is called {\em semilinear} if
all the languages in the family are semilinear.
Two words are said to be  {commutatively equivalent} if one is obtained from the other by rearranging the letters of the word. 
   Two languages $L_1$ and $L_2$ are said to be \emph{commutatively equivalent} if there exists a bijection $f\colon L_1 \to L_2$ such that every word $u\in L_1$ is commutatively equivalent to $f(u)$.
A language which is commutatively equivalent to a regular one will be called \emph{commutatively regular} \cite{di2013-code,di2013-diof,di2013-sem,CD-DLT-2018}. 
The counting function $f_L(n)$ of a language $L$ is the number of words of length $n$ in $L$. A language is counting regular if it has the same counting function as some regular language.

A {\em one-way $k$-counter machine} \cite{Ibarra1978} is a tuple
$M = (k,Q,\Sigma, \lhd,\delta, q_0, F)$, where
$Q, \Sigma, \lhd, q_0,F$ are respectively the finite set of states,
input alphabet, right input end-marker, initial state (in $Q$),
and accepting states (a subset of $Q$).
The transition function $\delta$ is a function from
$Q \times (\Sigma \cup \{\lhd, \lambda\}) \times \{0,1\}^k$
to the powerset of $Q \times \{-1,0, +1\}^k$, such that if
$\delta(q,a, c_1, \ldots, c_k)$ contains
$(p,d_1, \ldots, d_k)$ and $c_i = 0$ for some $i$, then $d_i \geq 0$
(to prevent negative values in any counter).
Then $M$ is deterministic if
$|\delta(q, a, i_1, \ldots, i_k) \cup \delta(q, \lambda, i_1, \ldots, i_k)| \leq 1$, for all $q \in Q, a \in \Sigma \cup \{\lhd\}, (i_1, \ldots, i_k) \in \{0,1\}^k$.
A configuration of $M$ is a ($k+2$)-tuple $(q,w, c_1, \ldots, c_k)$
representing that $M$ is in state $q$,
$w \in \Sigma^* \lhd \cup \{\lambda\}$ is still
to be read as input, and $c_1, \ldots, c_k \in \mathbb{N}_0$ are the contents of the $k$ counters. The derivation relation 
$\vdash_M$ is defined between configurations, whereby
$(q, aw, c_1, \ldots, c_k) \vdash_M (p, w, c_1 + d_1, \ldots, c_k + d_k)$,
if $(p,d_1, \ldots, d_k) \in \delta(q, a, \pi(c_1), \ldots, \pi(c_k))$.
Let $\vdash^*_M$ be the reflexive, transitive closure of $\vdash_M$.
A word $w \in \Sigma^*$ is accepted by $M$ if $(q_0, w\lhd, 0, \ldots, 0) \vdash_M^* (q, \lambda, c_1, \ldots, c_k)$, for some $q \in F, c_1, \ldots, c_k \in \mathbb{N}_0$. Furthermore, $M$ is $l$-reversal-bounded if it operates in such a way that in every accepting computation, the count on each counter alternates between non-decreasing and non-increasing and vice versa at most $l$ times.
The class of $k$-counter $l$-reversal-bounded machines is denoted by $\NCM(k,l)$, $\NCM(k)$ is the class of reversal-bounded $k$ counter machines, and $\NCM$ is the class of all reversal-bounded multicounter machines. Similarly, the deterministic variant is
denoted by $\DCM(k,l)$, $\DCM(k)$, and $\DCM$.


A context-free matrix grammar \cite{DassowPaun} (which we will henceforth simply call matrix grammar) is a tuple $G=(N,\Sigma,M,S)$, where $N$, $\Sigma$, and $M$ are respectively the finite sets of nonterminals, terminals, and matrix rules, and $S \in N$ is the start symbol.
We denote by $V$ the \emph{vocabulary} $V=N\cup \Sigma$ of $G$.
Each matrix rule $m \in M$ is a finite sequence $m = (p_1, \ldots, p_s)$ where each $p_i$, $1 \leq i \leq s$, is
a context-free production from $N$ to $V^*$. 
For $m\in M$, we define $x \derivestep[m]y, x,y \in V^*$ if $x= x_0 \derivestep[p_1]x_1
\derivestep[p_2] \cdots \derivestep[p_s] x_s = y$, where $\derivestep[p_i]$ is the standard context-free derivation relation.
Let $M^*$ be the set of the finite sequences of matrix rules and $\alpha \in M^*$.
If $\alpha = \lambda$, then $x \Rightarrow_{\alpha} x, x \in V^*$; if one has
\[
\alpha=m_1m_2\cdots 	m_s
\mbox{\quad and\quad}
x_0\derivestep[m_1]x_1\derivestep[m_2]\cdots \ x_{s-1}\derivestep[m_s]x_s,
\]
with $x_i\in V^*$, $m_j\in M$, $1\leq j\leq s$, $0\leq i\leq s$, then we write $x_0\derivestep[\alpha]x_s$.
The set of \emph{sentential forms} of $G$, $S(G) = \{w \in V^* \mid S \derivestep[\alpha] w, \alpha \in M^*\}$, and the \emph{language generated} by $G$, $L(G) = S(G) \cap \Sigma^*$.
We say that $G$ is ambiguous of degree $r$, where $r$ is a positive integer, if every word in $L(G)$ has at most $r$ distinct derivations in $G$, and some word in $L(G)$ has exactly $r$ distinct derivations. We define $G$ to be ambiguous of degree $\infty$ if there is no such $r$. If $G$ is ambiguous of degree $1$, then $G$ is said to be unambiguous. Let $\amb(G)$ be the degree of ambiguity of $G$.

For all $x\in V^*$, we denote by $\pi_N(x)$ (resp., $\pi_{\Sigma}(x)$) the word obtained deleting all terminals (resp., nonterminals) in $x$.
Notice that $\pi_N$ (resp., $\pi_{\Sigma}$) is a morphism of $V^*$ onto $N^*$ (resp., $\Sigma^*$).

%
%
%
%
%
Let $G$ be a matrix grammar.
The {\em index} of a derivation  
$x_0 \Rightarrow x_1 \Rightarrow \cdots \Rightarrow x_s$ of $G$ is the number
$\max_{0\leq i\leq s}|x_i|_N$.
The {\em index} of a word $w$ generated by $G$ is the minimal index of the derivations of $w$ in $G$.
The {\em index} of the grammar $G$ is the maximum index of the words $w\in L(G)$, provided that such a maximum exists.
In the opposite case, $G$ is said to have infinite index.
The {\em index} of a language $L$ is the minimal index of the grammars generating $L$.
Finally,   a language $L$ is said to be {\em  of finite index} if its index    is   finite.

The class of context-free matrix grammars is $\M$, and finite-index matrix grammars is 
$\M\fin$.

%

An $\ETOL$ system \cite{RozenbergFiniteIndexETOL} is a tuple $G = (V,{\cal P}, S, \Sigma)$,
where $V$ is a finite alphabet, $\Sigma\subseteq V$ is
the terminal alphabet, $S\in V$ is the axiom, and
${\cal P}$ is a finite set of production tables, where each 
$P \in {\cal P}$ is a
finite binary relation in $V \times V^*$. It is typically
assumed that for all production tables $P$ and each variable
$X \in V$, $(X,\alpha) \in P$ for some $\alpha \in V^*$.
If $(X,\alpha) \in P$, then we usually write $X \rightarrow_P \alpha$.
Elements of $V - \Sigma$ are called nonterminals.

Let $x = a_1 a_2 \cdots a_m, a_i \in V, 1 \leq i \leq m$, and let
$y \in V^*$. Then $x \Rightarrow_G y$, if there is a $P \in {\cal P}$
such that $y = \alpha_1 \cdots \alpha_m$ where 
$a_i \rightarrow \alpha_i \in P, 1 \leq i \leq m$. In this case, we also write $x \Rightarrow_P y$. Then $\Rightarrow_G^*$
is the reflexive, transitive closure of $\Rightarrow_G$, and the language generated by $G$, $L(G) = \{ x \in \Sigma^* \mid S \Rightarrow_G^* x\}$.
A letter $X \in V$ is {\em active} if there exists a table $P \in {\cal P}$ and a word $\alpha \in V^*$ such that $X \rightarrow_P \alpha$
and $\alpha \neq X$.
Then $A_G$ are the active symbols of $G$.
The system $G$ is in {\em active normal form} if
$A_G = V - \Sigma$. It is known that, given any $\ETOL$ system, another system $G'$ can be constructed in active normal form that generates the same language \cite{RozenbergFiniteIndexETOL}.
The index of a derivation $x_0 \Rightarrow_G x_1 \Rightarrow_G \cdots \Rightarrow_G x_s$ is
$\max_{0 \leq i \leq s} |x_i|_{A_G}$. The index of a word, grammar, and language are defined
just like for matrix grammars.

The system $G$ is said to be \emph{unambiguous} if, for all $w\in L(G)$, there is a unique derivation tree of $w$ in $G$.
This concept has been defined for $\EOL$ systems 
 in \cite{AmbiguityE0L}, but to our knowledge, not for $\ETOL$ systems generally. However, we define it here identically in Section \ref{sec:Finite-Index-ET0L-amb}.
An $\ETOL$ system is called {\em deterministic} ($\EDTOL$) if each table $P \in {\cal P}$ satisfies the following condition: 
if $X \rightarrow_P \alpha$ and $X \rightarrow_P \beta,$ then $\alpha = \beta$.
An $\ETOL$ system is said to be a  {\em $\EOL$  system (resp., {\em $\EDOL$  system) if  the system
(resp., deterministic system) has a sole production table. 
The class of all $\ETOL$ systems is denoted by
$\ETOL$, and similarly for other types of L-systems.

 For a class of machines ${\cal M}$, we use the notation $\LL({\cal M})$ to denote the family of languages accepted by machines in ${\cal M}$. For a class of grammars ${\cal G}$, $\LL({\cal G})$ denotes the family of languages generated by the grammars.
A language family $\LL$ is a {\em trio} \cite{G75} if it is closed under
$\lambda$-free morphism, inverse morphism, and intersection
with regular languages. 


\section{Bounded Languages and Counter Machines}\label{sec:Boun-Lang-Count-Machines}

In this section, we will define four different notions for describing languages that are both bounded and semilinear.
First, a function is defined.
Given words $w_1, \ldots, w_k \in \Sigma^+$, define a function $\phi$ from $\mathbb{N}_0^k$ to $\Sigma^*$ that maps $\phi(i_1, \ldots, i_k)$ to $w_1^{i_1} \cdots w_k^{i_k}$, which is extended in the natural way from subsets of $\mathbb{N}_0^k$ to $\Sigma^*$.

\begin{definition}
Let $\Sigma =  \{a_1, \ldots , a_n \}$,  $w_1, \ldots, w_k \in \Sigma^+$, and $Q_1  \subseteq  \natzero^k$  and  $Q_2  \subseteq  \natzero^n$ 
be semilinear sets. 
\begin{enumerate}
\item
If  $L = \phi(Q_1)$,
then $L$ is called the {\em bounded Ginsburg semilinear language}
{\em induced} by $Q_1$.
\item
If $L =  \{ w  \mid  w =  w_1^{i_1} \cdots w_k^{i_k}, i_1, \ldots, i_k \in \mathbb{N}_0, \psi(w) \in  Q_2 \}$,
then $L$ is called the {\em bounded Parikh semilinear language}
{\em induced} by $Q_2$.
\item
If $L = \{ w \mid w =  \phi(i_1, \ldots, i_k), (i_1, \ldots , i_k) \in  Q_1,
\psi(w) \in Q_2 \}$, 
then $L$ is called the {\em bounded Ginsburg-Parikh semilinear language}
induced by $Q_1$ and $Q_2$.
\item 
If $L \subseteq w_1^* \cdots w_k^*$, and $\psi(L) = Q_2$,
then $L$ is called a {\em bounded general semilinear language}
with Parikh image of $Q_2$.
\end{enumerate}
\label{boundedDefs}
\end{definition}
Traditionally, bounded Ginsburg semilinear languages are referred to as simply bounded semilinear languages \cite{IbarraSeki}. However, in this paper, we will use the term bounded Ginsburg semilinear language to disambiguate with other types. 
Note that a bounded Parikh semilinear language is a special case
of bounded general semilinear language.

\noindent
\begin{example}  Consider the following languages:
\begin{itemize}
\item
Let $L_1   =  \{ w \mid w = (abb)^i  (bab)^j  (abb)^k, 0 < i < j < k \}$.
Here, with the semilinear set $Q_1 = \{(i, j, k) ~|~    0 < i < j < k\}$,
then $L_1 = \{ w \mid w = (abb)^i (bab)^j (abb)^k, (i,j,k) \in Q_1\}$,
and therefore $L_1$ is bounded Ginsburg semilinear.
\item
Let $L_2 = \{ w \mid  w = (abb)^i (aba)^j , i,j >0, 0 <|w|_a = |w_b| \}$.
Here, using the semilinear set $Q_2 = \{(r, r) ~|~  0 < r \}$, it can be seen that $L_2$ is bounded Parikh semilinear.
\item
Let $L_3 = \{ w \mid  w = (abbb)^i (aab)^j , 0 < i < j, 0 <|w|_a < |w_b| \}$.
Using $Q_1 = \{(r, s) ~|~ 0 < r < s \}$, and
$Q_2 = \{(r, s) \mid  0 < s < r \}$, then 
$L_3$ is bounded Ginsburg-Parikh semilinear. Hence, both $Q_1$
and $Q_2$ help define $L_3$. For example, if $i =2, j = 3$, then
$w = (abbb)^2(aab)^3 \notin L_3$ despite $2 < 3$, since $|w|_a = 8 < |w|_b = 9$.
But if $i=2, j = 5$, then $w = (abbb)^2(aab)^5 \in L_3$ since
$2<5$ and $|w|_a = 12 > |w|_b = 11$.
\item
Let $L_4 = \{a^{2^i} b \mid i >0\} \cup \{ba^i \mid i >0\}$. Then
$L_4$ is bounded as it is a subset of $a^*b^*a^*$, and has the same Parikh image as the
regular language $\{ba^i \mid i >0\}$ and is therefore semilinear, and hence bounded general semilinear. It will become
evident from the results in this paper that $L_4$ is not
bounded Ginsburg-Parikh semilinear.
\end{itemize}
\end{example}
Note that given the semilinear sets and the words
$w_1, \ldots, w_k$ in Definition \ref{boundedDefs},
there is only one bounded Ginsburg,
bounded Parikh, and bounded Ginsburg-Parikh semilinear language
induced by the semilinear sets. But for bounded general
semilinear languages, this is not the case, as $L_4$ in the
example above has the same Parikh image as the regular language
$\{ba^i \mid i >0\}$.

The following known results are required:

\begin{proposition} \label{thm1}
Let $\Sigma = \{a_1, \ldots , a_n\}$ and $w_1, \ldots , w_k \in \Sigma^+$.
\begin{enumerate}
\item \cite{Ibarra1978} If $L \subseteq w_1^* \cdots w_k^*$ is in 
$\LL(\NCM)$, then $Q_L = \{(i_1, \ldots , i_k) \mid w_1^{i_1} \cdots w_k^{i_k} \in L\}$
is a semilinear set (i.e.\ every bounded language in $\LL(\NCM)$ is bounded Ginsburg semilinear).

\item \cite{IbarraSeki}
If $Q \subseteq \mathbb{N}_0^k$ is a semilinear set, then 
$\phi(Q)
\in \LL(\NCM)$ (every bounded Ginsburg semilinear language is in $\LL(\NCM)$).
\item
\cite{Ibarra1978} If $L \subseteq \Sigma^*$ is in $\LL(\NCM)$,
then $\psi(L)$ is a semilinear set. 
\end{enumerate}
\end{proposition}

\begin{proposition} \cite{IbarraSeki} \label{thm2}
$\LL(\NCM)\bd = \LL(\DCM)\bd$.
\end{proposition}

\begin{corollary} \label{cor1}
In Proposition \ref{thm1}, $\NCM$ can be replaced by $\DCM$.
\end{corollary}

The following lemma is also required, which
is generally known (e.g., it can be derived from
the results in \cite{eDCM}), but a short proof is given for
completeness.
\begin{lemma} \label{new}
Let $\Sigma = \{a_1, \ldots, a_n \}$.
If $Q \subseteq  \mathbb{N}_0^n$ is a semilinear set,
then $A = \{ w ~|~ w \in \Sigma^*, \psi(w) \in Q \} \in \LL(\NCM)$.
\end{lemma}
\begin{proof}
Since $\LL(\NCM)$ is closed under union, it is sufficient
to prove the result for the case when $Q$ is a linear
set.  Let $Q = \{ v \mid v = \vec{v_0} + i_1\vec{v_1} + \cdots + i_r\vec{v_r},$
each $i_j \in  \mathbb{N}\}$,
where $\vec{v_0} = (v_{01}, \ldots, v_{0n})$ is the constant
and $\vec{v_j} = (v_{j1}, \ldots, v_{jn})$ ($1 \le j \le r$)
are the periods.  Construct an $\NCM$ $M$ with
counters $C_1, \ldots, C_n$ which, when given input $w \in \Sigma^*$,
operates as follows:
\begin{enumerate}
\item
$M$ reads $w$ and stores $|w|_{a_i}$ in
counter $C_i$ ($1 \le i \le n$).
\item
On $\lambda$-moves, $M$ decrements $C_i$ by $v_{0i}$
for each $i$ ($1 \le i \le n$).
\item
For each $1 \le j \le r$, on $\lambda$-moves, $M$ decrements each
$C_i$ ($1 \le i \le n$) by $v_{ji}$ repeatedly some nondeterministically guessed number of times; this has the effect of decreasing by  $k_jv_{ji}$, for some $k_j \geq 0$.
\item
$M$ accepts when all counters are zero.
\end{enumerate}
\noindent
Then, $L(M) = A$.
\qed
\end{proof}



Next, the relationship is examined between
bounded Ginsburg semilinear languages, bounded Parikh semilinear languages, 
bounded Ginsburg-Parikh semilinear languages, and bounded general semilinear languages.
To start, the following proposition is needed:
%

\begin{proposition} \label{notrecursive}
Let $\LL$ be any family of languages which  is contained in the family
of recursively enumerable languages.  Then there is a bounded general semilinear
language that is not in $\LL$.
\end{proposition}
\begin{proof}
Take any non-recursively enumerable language  $ L \subseteq a^*$.
Let $b, c$ be new symbols, and consider
$L' = b L c \cup c a^* b$.  Then $L'$ is bounded, since it is a
subset of $b^* a^* c^* a^* b^*$.
Clearly, $L'$ has the same Parikh image as the regular language
$ca^*b$. Hence, $\psi(L') = \{ (i, 1, 1) ~|~  i \ge 0 \}$, which is semilinear.
But $L'$ cannot be recursively enumerable, otherwise by intersecting it
with the regular language $b a^* c$,  $bLc$ would also be recursively
enumerable.
But $bLc$ is  recursively enumerable
if and only if $L$ is recursively enumerable. This is a 
contradiction, since $L$ is not recursively enumerable.
Thus, $L' \notin \LL$.
\qed
\end{proof}
Therefore, there are bounded general semilinear languages that
are not recursively enumerable. 

Next, the comparison between the four types of languages is made.

\begin{proposition} \label{BoundedRelationship}
The family of bounded Parikh semilinear
languages is a proper subset of the 
family of bounded Ginsburg semilinear languages, 
which is equal to the family of bounded Ginsburg-Parikh semilinear languages, which is a proper subset of
the family of bounded general semilinear languages.
\end{proposition}
\begin{proof}
First note that every bounded Ginsburg semilinear language is a
bounded Ginsburg-Parikh semilinear language by
setting $Q_2 = \natzero^n$ (in Definition \ref{boundedDefs}). Also,
every bounded Parikh
semilinear language is a bounded Ginsburg-Parikh semilinear language 
by setting $Q_1 = \natzero^k$.
So, both the families of bounded Ginsburg semilinear languages 
and bounded Parikh semilinear languages are a subset
of the bounded Ginsburg-Parikh semilinear languages.

Next, it will be shown that every bounded Parikh semilinear
language is a bounded Ginsburg semilinear language.
Let $L$ be a bounded Parikh semilinear language, induced
by semilinear set $Q$. Then $L = \{w \mid w = w_1^{i_1} \cdots w_k^{i_k}, i_1, \ldots, i_k \in \mathbb{N}_0,
(|w|_{a_1}, \ldots, |w|_{a_n}) \in Q\}$.
Let $A_Q = \{w ~|~ w \in \{a_1, \ldots, a_n\}^*, \psi(w) \in Q\}$.
Clearly, $L = A_Q \cap w_1^* \cdots w_k^*$.
By Lemma \ref{new}, $A_Q$ is in $\LL(\NCM)$ and since $\LL(\NCM)$ is closed
under intersection with regular sets \cite{Ibarra1978}, 
$L$ is also in $\LL(\NCM)$.
Then, by Proposition \ref{thm1} Part 1, $L$ is bounded Ginsburg semilinear.

Notice that the bounded Ginsburg-Parikh semilinear language induced by $Q_1, Q_2$ is the intersection of the bounded Ginsburg semilinear
set induced by $Q_1$, with the bounded Parikh semilinear language
induced by $Q_2$. From the proof above, every bounded Parikh
semilinear language is in fact a bounded Ginsburg semilinear language.
Hence, every bounded Ginsburg-Parikh semilinear language is the
intersection of two bounded Ginsburg semilinear languages.
As every bounded Ginsburg semilinear language is in $\LL(\NCM)$ by Proposition \ref{thm1} Part 2,
and $\LL(\NCM)$ is closed under intersection \cite{Ibarra1978}, it follows
that every bounded Ginsburg-Parikh semilinear language is in $\LL(\NCM)$.
By an application of Proposition \ref{thm1} Part 1 followed
by Part 2, every Ginsburg-Parikh semilinear language must therefore be 
a bounded Ginsburg semilinear language.

To show that bounded Parikh semilinear languages
are properly contained in bounded Ginsburg languages,
consider the bounded Ginsburg semilinear language
$L = \{a^i  b^i a^i  \mid i  >  0 \}$
induced by  semilinear set $Q_1 = \{ (i , i,  i) \mid i > 0 \}$.
Now the Parikh image of $L$ is the semilinear set
$Q_2 = \{(2i , i) ~|~ i > 0\}$. Thus, if the fixed
words are $a$, $b$, $a$ (whereby these are the words chosen
to define the bounded language), then the bounded
Parikh semilinear language induced by
$Q_2$ is  $L' = \{a^i b^k a^j \mid  i+j = 2k  > 0 \}$,
which is different from $L$. It is clear that this is true
for all fixed words.

It suffices to show that the family of bounded Ginsburg semilinear languages is strictly contained
in the family of bounded general semilinear languages. Containment can be seen as follows:
Let $L$ be a bounded Ginsburg language. Every bounded Ginsburg language is in $\LL(\NCM)$
by Proposition \ref{thm1} Part 2, and all $\LL(\NCM)$ languages are semilinear by Proposition \ref{thm1} Part 3. 
Thus $L$ is semilinear, and $L$ is also bounded. Hence, $L$ is bounded general semilinear.
Strictness follows from Proposition \ref{notrecursive} and
the fact that  all $\LL(\NCM)$ languages are recursive
\cite{Ibarra1978}.
\qed
\end{proof}

In fact, as long as a language family contains a simpler
subset of bounded Ginsburg semilinear languages, and is
closed under $\lambda$-free morphism, then it is enough to
imply they contain all bounded Ginsburg semilinear
languages.
\begin{proposition} \label{thm3}
Let ${\cal L}$ be a language family that contains all
distinct-letter-bounded Ginsburg semilinear languages
and is closed under $\lambda$-free morphism. Then ${\cal L}$
contains all bounded Ginsburg semilinear languages.
\end{proposition}
\begin{proof}

Let $w_1, \ldots, w_k \in \Sigma^+$, and let 
$L\subseteq w_1^* \cdots w_k^*$ be a bounded Ginsburg
semilinear language induced by $Q_1$.
Let $b_1, \ldots, b_k$ be new distinct symbols. 
Consider the bounded Ginsburg semilinear 
language $L' \subseteq b_1^* \cdots b_k^*$ induced by 
$Q_1$. Then $L' \in {\cal L}$ by assumption.
Finally, apply morphism $h$ on $L'$ defined by
$h(b_i) = w_i$  for each $i$.
Then $h(L') = L$, which must be in ${\cal L}$, since ${\cal L}$ is
closed under $\lambda$-free morphism.
\qed
\end{proof}

Furthermore, as long as a language family is a semilinear
trio, all bounded languages in the family are bounded
Ginsburg semilinear languages.
\begin{proposition} \label{thm4}
Let $\Sigma = \{a_1, \ldots, a_n\}$, $w_1, \ldots, w_k \in \Sigma^+$,
${\cal L}$ is a semilinear trio, and let
$L  \subseteq w_1^* \cdots w_k^*, L \in {\cal L}$. There is a semilinear set
$Q_1$ such that $L$ is the bounded Ginsburg semilinear language
induced by $Q_1$.
\end{proposition}
\begin{proof}
Let $b_1, \ldots, b_k$ be new distinct symbols, and
$L_1 = \{b_1^{i_1} \cdots b_k^{i_k} \mid w_1^{i_1} \cdots w_k^{i_k} \in L \}$.
Then, since ${\cal L}$ is closed under $\lambda$-free finite 
transductions (every trio is closed under $\lambda$-free finite transductions \cite{G75}, Corollary 2 of Theorem 3.2.1), $L_1 \in {\cal L}$,
as a transducer can read $w_1$, and output $b_1$ some number
of times (nondeterministically chosen), followed by $w_2$, etc.
This transducer is $\lambda$-free as these can read a fixed word from the input and output a letter.
Let $Q_1$ be the Parikh image of $L_1$, which is semilinear
by assumption. It follows that $L$ is the bounded
Ginsburg semilinear language induced by $Q_1$.
\qed
\end{proof}
Hence, all bounded languages in semilinear
trios are ``well-behaved'' in the sense that they are bounded Ginsburg semilinear.
For these families, 
bounded languages, and bounded Ginsburg semilinear languages
coincide.
\begin{corollary}
\label{allGinsburg}
Let ${\cal L}$ be a semilinear trio. Then $L \in {\cal L}$
is bounded if and only if $L$ is bounded Ginsburg semilinear.
Hence, ${\cal L}\bd= \{L \mid L \in {\cal L}$ is bounded
Ginsburg semilinear$\}$.
\end{corollary}
Note that this is not necessarily the case for non-semilinear trios.
For example, the language family $\LL(\ETOL)$ contains the non-semilinear language
$\{a^{2^n} \mid n>0\}$ which is bounded but not semilinear \cite{RS}. Hence, $\LL(\ETOL)$
contains languages that are bounded general semilinear, $\{a^{2^n} \mid n>0\}b \cup ba^*$,
but not bounded Ginsburg
semilinear in a similar fashion to Proposition \ref{notrecursive}. But this cannot
happen within semilinear trios.

Also, since all bounded Ginsburg semilinear languages are commutatively regular \cite{di2013-code,di2013-diof,di2013-sem},
we obtain the following corollary.
\begin{corollary}
Let $\LL$ be a semilinear trio. All bounded languages in $\LL$ are commutatively regular.
\end{corollary}

Moreover, for an arbitrary semilinear trio ${\cal L}$, it is possible
to compare all bounded languages in ${\cal L}$ to the set of all
bounded Ginsburg semilinear languages, which are exactly the
bounded languages in
$\LL(\NCM)$.
\begin{proposition}
\label{ifandonlyif}
Let $\LL$ be a semilinear trio.
Then $\LL\bd \subseteq \LL(\NCM)\bd = \LL(\DCM)\bd$ and the
following conditions are equivalent:
\begin{enumerate}
\item $\LL\bd = \LL(\NCM)\bd = \LL(\DCM)\bd$.
\item $\LL$ contains all
bounded Ginsburg semilinear languages.
\item $\LL$ contains all bounded Parikh semilinear languages.
\item $\LL$ contains all distinct-letter-bounded Ginsburg semilinear languages. 
\end{enumerate}
\end{proposition}
\begin{proof} 
$\LL(\NCM)\bd = \LL(\DCM)\bd$ follows from Proposition \ref{thm2}.

Also, all distinct-letter bounded Ginsburg semilinear languages are bounded Parikh semilinear, and all bounded Parikh semilinear languages are bounded 
Ginsburg semilinear languages by Proposition \ref{BoundedRelationship}, and thus 2 implies 3 and 3 implies
4. The other direction 
follows from Proposition \ref{thm3}, and thus 4 implies 3 and 3 implies
2. Hence, 2, 3, and 4 are equivalent.

Consider any bounded language $L\subseteq w_1^* \ldots w_k^* \in {\cal L}$. Then there is a semilinear set $Q$ such that $L$ is the bounded
Ginsburg semilinear language induced by $Q$, by Corollary \ref{allGinsburg}.
By Proposition \ref{thm1} Part 2, $L \in \LL(\NCM)$. 
Hence, $\LL\bd \subseteq \LL(\NCM)\bd$.

If ${\cal L}$ does not contain all distinct-letter-bounded Ginsburg
semilinear
languages, then
$\LL\bd \subsetneq \LL(\NCM)\bd$, as $\LL(\NCM)$ does, by Proposition 
\ref{thm1} Part 2. Otherwise, if ${\cal L}$ does contain all distinct-letter-bounded Ginsburg semilinear languages,
then it contains all bounded Ginsburg semilinear languages
by Proposition \ref{thm3}, and
then by Proposition \ref{thm1}, 
all bounded languages in $\LL(\NCM)$
are in ${\cal L}$. Hence, 4 is equivalent to 1.
\qed
\end{proof}

Since this result was shown, the latter two authors provided a characterization of the smallest full trio containing the bounded Ginsburg semilinear languages by using restrictions of $\NCM$ \cite{InstructionsTCS}.

As a consequence of Proposition \ref{ifandonlyif}, every bounded language in any semilinear trio $\LL$ is in
$\LL(\DCM)$. The next proposition shows that if
the trio properties are effective, and the family is effectively semilinear (which means, there is an algorithm which takes as input a finite representation of a member of the family, and it determines the constant and periods of each of the linear sets), there is an algorithm to effectively construct a $\DCM$ machine accepting a given bounded language in $\LL$.
\begin{proposition}
Let ${\cal L}$ be any language family that is effectively closed
under the trio operations, and is effectively semilinear.
For each bounded language $L \in {\cal L}$, 
$L \subseteq w_1^* \cdots w_k^*$ ($w_1, \ldots, w_k$ are given),
it is possible to build a $\DCM$ machine accepting $L$.
\label{toDCM}
\end{proposition}
\begin{proof}
Following the proof that $\LL\bd \subseteq \LL(\NCM)\bd$, given $L\subseteq w_1^* \cdots w_k^* \in \LL$, 
Corollary \ref{allGinsburg} indicates that there is a semilinear set $Q$ such that $L = \phi(Q)$.
Examining the proof of Proposition \ref{thm4} (used for Corollary \ref{allGinsburg}), the transducer can
be built if $w_1, \ldots, w_k$ are known. Given a transducer, it is possible to construct a sequence of trio operations
simulating it (Corollary 2 of Theorem 3.2.1 in \cite{G75}).
Since semilinearity is effective in $\LL$, it is possible to construct the constant and periods of each linear set in the proof
of Proposition \ref{thm4}. In Proposition \ref{thm1} part 2, construction of $\phi(Q)$  from $Q$ in \cite{IbarraSeki}
is effective given $w_1, \ldots, w_k$. Lastly, the construction of a $\DCM$ from an $\NCM$ in \cite{IbarraSeki} is effective.
\qed
\end{proof}
This provides a deterministic machine model to accept
all bounded languages from these language families defined by nondeterministic machines and grammars. Moreover, $\DCM$ machines have
many decidable properties, allowing for algorithms to be used
on them.
\begin{corollary} Let ${\cal L}_1$ and ${\cal L}_2$ be two language families effectively closed under the trio operations, and effectively semilinear. 
It is decidable, for $L_1 \in {\cal L}_1\bd$, and 
$L_2 \in {\cal L}_2\bd$, whether $L_1 \subseteq L_2$, whether $L_1 = L_2$ (and the words over which $L_1$ and $L_2$ are bounded are given), and whether $L_1 \cap L_2 \neq \emptyset$.
\end{corollary}
\begin{proof}
This follows since every bounded language within both language
families are in $\LL(\DCM)$ (effectively) by Proposition \ref{toDCM},
and containment, equality, and disjointness are decidable for
$\LL(\DCM)$ \cite{Ibarra1978}.
\qed
\end{proof}

Hence, the results of this section together show that it is not needed to devise separate proofs for the standard decision problems applied to bounded languages in semilinear trios. All of the standard decision problems are always decidable, and their decidability even extends to testing containment, equivalence, and disjointness between languages from different families, e.g.\ one language generated by a finite-index $\ETOL$ system \cite{RozenbergFiniteIndexETOL}, and one by a multi-pushdown machine \cite{multipushdown}, created in entirely different ways.

\section{Finite-Index $\ETOL$ and Finite-Index Matrix Languages}
\label{Finite-Index-BounLan}

It is known that the family of finite-index $\ETOL$ languages is a semilinear
trio \cite{RozenbergFiniteIndexETOL}, and therefore
all bounded languages in it are $\DCM$ languages, by
Proposition \ref{ifandonlyif}. We will show that the bounded languages
in the two families are identical. This demonstrates an application of Proposition \ref{ifandonlyif}.
\begin{lemma} \label{lem1}
Let $a_1, \ldots, a_k$ be distinct symbols, and $Q  \subseteq \natzero^k$
be a semilinear set. Then $L = \{a_1^{i_1} \cdots a_k^{i_k}  \mid (i_1, \ldots, i_k) \in Q\} \in \LL(\ETOL\fin)$.
\end{lemma}
\begin{proof}
Let $L$ be a letter-bounded Ginsburg 
semilinear language of the form above, and let 
$\Sigma = \{a_1, \ldots, a_k\}$. Then $\psi(L)$ is a finite union 
of linear sets. Consider each of the linear sets, $Q$, where there are
$\vec{v_0}, \vec{v_1}, \ldots, \vec{v_r} \in \natzero^k$ ($\vec{v_0}$ the constant, the rest the periods) with
$Q = \{ \vec{v_0} + i_1\vec{v_1} + \cdots + i_r\vec{v_r} \mid i_1, \ldots, i_r \in \natzero\}$. Assume that $r \geq 1$, otherwise
the set is finite, where the case is obvious.

We create an $\ETOL$ system $G_Q = (V,{\cal P}, S, \Sigma)$ 
as follows:
${\cal P} = \{P_0, P_1\}$, $V =\Sigma \cup \{ Z\} \cup 
\{X_{i,j} \mid 1 \leq i \leq k, 1 \leq j \leq r\}$  ($Z$ is a ``dead'' nonterminal),
and the productions are:
\begin{enumerate}
\item Add $S \rightarrow_{P_1} 
a_1^{\vec{v_0}(1)} X_{1,1} a_2^{\vec{v_0}(2)} X_{2,1} \cdots a_k^{\vec{v_0}(k)} X_{k,1}$ and $S \rightarrow_{P_0} Z$.
 
\item For all $X_{i,j} \in V$, add 
$X_{i,j} \rightarrow_{P_0} a_i^{\vec{v_j}(i)} X_{i,j}$.

\item For all $X_{i,j} \in V, 1 \leq j < r$, add 
$X_{i,j} \rightarrow_{P_1} X_{i,j+1}$.

\item For all $X_{i,r} \in V$, add $X_{i,r} \rightarrow_{P_1}
\lambda$.

\item $a \rightarrow_P a$ is a production for every
$a \in \Sigma \cup \{Z\}$, and $P \in {\cal P}$.

\end{enumerate}

\begin{claim}
$L(G_Q) = \{a_1^{l_1} \cdots a_k^{l_k}  \mid (l_1, \ldots, l_k) \in Q\}$,
and $G_Q$ is of index $k$.
\end{claim}
\begin{proof}
``$\subseteq$''
Let $w \in L(G_Q)$. Thus, there exists $ S \Rightarrow_{Q_1} x_1
\Rightarrow_{Q_2} \cdots \Rightarrow_{Q_s} x_s =w \in \Sigma^*$,
$Q_l \in \{P_0, P_1\}$, $1 \leq l \leq s$. Then 
$Q_1 Q_2 \cdots Q_s$ must be of the form 
$$P_1 P_0^{i_1} P_1 P_0^{i_2} P_1 \cdots P_0^{i_r} P_1,$$
where $i_j \in \natzero$, by the construction.
We will show by induction that, for all $0 \leq j < r$,
$x_{i_1 + \cdots + i_j + j +1}$ (this is the sentential
form after the ($j+1$)st application of the production table $P_1$)
 is equal to
\begin{equation}
a_1^{\vec{v_0}(1) + i_1 \vec{v_1}(1)+ \cdots + i_j\vec{v_j}(1) }  X_{1,j+1} a_2^{\vec{v_0}(2)+ i_1 \vec{v_1}(2)+ \cdots + i_j\vec{v_j}(2)} X_{2,j+1} \cdots a_k^{\vec{v_0}(k) + i_1 \vec{v_1}(k) + \cdots + i_j\vec{v_j}(k)} X_{k,j+1},
\label{semilinearword-a}
\end{equation} and
for $j = r$, it is 
$$a_1^{\vec{v_0}(1) + i_1 \vec{v_1}(1)+ \cdots + i_r\vec{v_r}(1) } a_2^{\vec{v_0}(2)+ i_1 \vec{v_1}(2)+ \cdots + i_r\vec{v_r}(2)} \cdots a_k^{\vec{v_0}(k) + i_1 \vec{v_1}(k) + \cdots + i_r\vec{v_r}(k)}.$$
The base case, $j =0$, follows since 
$x_1 = a_1^{\vec{v_0}(1)} X_{1,1} a_2^{\vec{v_0}(2)} X_{2,1} \cdots a_k^{\vec{v_0}(k)} X_{k,1}$ using the production of type 1.

Let $0 \leq j <r$ and assume that $x_{i_1 + \cdots + i_j + j+1}$ is
equal to the string in Equation (\ref{semilinearword-a}).
Then, productions created in step 2 must get applied $i_{j+1}$ times, followed by one application created in step 3 if $j+1<r$, or
one application created in step 4 if $j+1 = r$. Then 
it is clear that the statement holds for $j+1$ as well.

It is also immediate that every sentential form in $G_Q$ has
at most $k$ active symbols, and therefore it is of index $k$.

``$\supseteq$''
Let $w = a_1^{l_1} \cdots a_k^{l_k}$, with $(l_1, \ldots, l_k) \in Q$.
Then $(l_1, \ldots, l_k) = \vec{v_0} + i_1 \vec{v_1} + \cdots
+ i_r \vec{v_r}$, for some $i_1, \ldots, i_r \in \natzero$.
Then, by applying a production table sequence of the form
$P_1 P_0^{i_1} P_1 \cdots P_0^{i_r} P_1$, this changes the derivation as
follows:
\begin{tabbing}
$S \Rightarrow  a_1^{\vec{v_0}(1)} X_{1,1} a_2^{\vec{v_0}(2)} X_{2,1} \cdots a_k^{\vec{v_0}(k)} X_{k,1} $\\[.15cm]
$\Rightarrow^* a_1^{\vec{v_0}(1) + i_1 \vec{v_1}(1)} X_{1,2} a_2^{\vec{v_0}(2)+ i_1 \vec{v_1}(2)} X_{2,2} \cdots a_k^{\vec{v_0}(k) + i_1 \vec{v_1}(k)} X_{k,2}$ \\[.15cm]
$\Rightarrow^* a_1^{\vec{v_0}(1) + i_1 \vec{v_1}(1)+ \cdots + i_r\vec{v_r}(1) }  a_2^{\vec{v_0}(2)+ i_1 \vec{v_1}(2)+ \cdots + i_r\vec{v_r}(2)} \cdots a_k^{\vec{v_0}(k) + i_1 \vec{v_1}(k) + \cdots + i_r\vec{v_r}(k)}$\\[.15cm]
$=  a_1^{l_1} \cdots a_k^{l_k}.$
\end{tabbing}
\end{proof}

Hence $G_Q$ can generate all strings in $\{a_1^{l_1} \cdots a_k^{l_k} \mid (l_1,\ldots, l_k) \in Q\}$. As $L$ is semilinear, it is
the finite union of linear sets. Thus, $L$ can be generated in this
manner since $k$-index $\ETOL$ is closed under union \cite{RozenbergFiniteIndexETOL}.
\qed
\end{proof}

Next, finite-index $\ETOL$ languages coincide with languages
accepted by other types of finite-index grammars, such as
$\EDTOL$, context-free programmed grammars (denoted by $\CFP$), ordered grammars (denoted by $\ORDER$),
and matrix grammars (with the `fin' subscript
used for each family)
\cite{RozenbergFiniteIndexGrammars}.
\begin{proposition} \label{thm5}
The bounded languages in the following families coincide,
\begin{itemize}
\item $\LL(\NCM)$,
\item $\LL(\DCM)$,
\item $\LL(\ETOL\fin) = \LL(\EDTOL\fin) = \LL(\CFP\fin) = \LL(\ORDER\fin) = \LL(\MATRIX\fin)$,
\item the family of bounded Ginsburg semilinear languages.
\end{itemize}
\end{proposition}
\begin{proof}
$\ETOL\fin$ coincides with languages generated
by all the other grammar systems of finite-index \cite{RozenbergFiniteIndexGrammars}, and so it follows that the bounded languages within each coincide as well.
The rest follows from Proposition \ref{ifandonlyif} and Lemma \ref{lem1}.
\qed
\end{proof}

From Proposition \ref{thm5}, we know the bounded languages within
 $\NCM$ and $\ETOL\fin$ coincide (which are strictly included
in the bounded languages within $\ETOL$ as the non-semilinear
language $\{a^{2^n} \mid n \geq 0\}$ is in $\ETOL$). Next, we will address the relationship between $\NCM$ and 
$\ETOL\fin$ (over non-bounded languages).

We observe that there are $\LL(\ETOL\fin)$ languages that
are not in $\LL(\NCM)$.
\begin{lemma} \label{strictness}
There exists a language $L \in \LL(\ETOL\fin) - \LL(\NCM)$.
\end{lemma}
\begin{proof}
Consider $L = \{ x \# x \mid x \in \{a,b\}^+ \}$.
It is easy to construct an $\ETOL$ system of finite index to
generate $L$.  We will show that $L$ cannot
be accepted by any $\NCM$.

It was shown in \cite{Baker1974} that for
any $\NCM$ $M$, there is a constant $c$ (which depends
only on $M$) such that if $w$ is accepted by $M$,
then $w$ is accepted by $M$ within $cn$ steps, where
$n = |w|$.  So suppose $L$ is accepted by $M$.
Consider a string $x \#x$, where $n = |x| \ge 1$.  Then
$M$'s input head will reach $\#$ within  $cn$
steps.  If $M$ has $k$ counters, the number of
configurations (state and counter values) when
$M$ reaches $\#$ is  $O(s(cn)^k)$, where $s$
is the number of states (as each counter
can grow to at most $cn$ in $cn$ moves).  Since there are $2^n$
strings of  the form $x \# x$, where $x \in \{a,b\}^+$
and $|x| = n$, it would follow that for large
enough $n$, there are distinct strings $x$ and $y$
of length  $n$ such that $x \#y$ would be accepted
by $M$.  This is a contradiction. Hence
$L$ cannot be accepted by any $\NCM$.
\qed
\end{proof}

%
%
%

It is an open problem whether there are
languages in $\LL(\NCM)$ that are not in $\ETOL\fin$.
We conjecture that over the alphabet
$\Sigma_k = \{a_1, \ldots, a_k\}$, the language 
$\LL_k = \{w \mid |w|_{a_1} = \cdots = |w|_{a_k}\}$ is not in
$\LL(\ETOL\fin)$.  
 A candidate witness language that we initially thought of is the
one-sided Dyck language on one letter which is not in $\ETOL\fin$
\cite{Rozoy}. However, this language
cannot be accepted by any blind counter machine, which is equivalent
to an $\NCM$ \cite{G78}.
It is worth noticing that in \cite{beau} it is shown that the all the families of
$2$-sided Dyck languages ${\cal D}_k ^*$ over $k$ generators, with $k\geq 2$, are not in $\ETOL\fin$, while
it is still open whether ${\cal D}_1 ^* = \LL_1$ is in  $\LL(\ETOL\fin)$.  

In the subsequent part of the paper, we will provide a partial answer in the affirmative to the latter conjecture. Precisely, we will
prove that an unambiguous finite-index $\ETOL$ system cannot generate ${\cal L}_2$. In order to prove this result, we need
to discuss the property of ambiguity for such structures. This will be done in the next section.


\section{Finite-Index $\ETOL$, Ambiguity, and Connections to Matrix Grammars}\label{sec:Finite-Index-ET0L-amb}

Next, the concept of ambiguity will be defined.
To our knowledge, ambiguity has never been defined generally for $\ETOL$. However, it has been studied in \cite{AmbiguityE0L} for $\EOL$ systems, which are a special case of $\ETOL$ with only one table. We will use the same definition as their paper.

\begin{definition}
Let $G$ be an $\ETOL$ system. We say that $G$ is {\em ambiguous of degree $r$}, where $r$ is a positive integer, if every word in $L(G)$ has at most $r$ distinct derivation trees in G; and some word in $L(G)$ has exactly $r$ distinct derivation trees. We define $G$ to be {\em ambiguous of degree} $\infty$ if there is no such $r$. If $G$ is ambiguous	 of degree $1$, then $G$ is said to be {\em unambiguous}. Let $\amb(G)$ be the degree of ambiguity of $G$.
\end{definition}

\begin{remark}
Notice that every $\ETOL$ system $G$ such that $L(G) \neq \emptyset$ in active normal form has $\amb(G) = \infty$, because the terminals can rewrite to themselves arbitrarily many times, and therefore there are always an infinite number of distinct derivation trees for every word in $L(G)$. But we will see that this is not always the case for $\ETOL$ systems not in active normal form. 
\label{neverambiguous}
\end{remark}

It is possible to interpret terminals of $\ETOL$ systems in a similar manner to context-free grammars, where they do not get rewritten, and they are not members of $V$. We define {\em reduced $\ETOL$ systems} $G = (V,{\cal P}, S, \Sigma)$ to be as in the definition of $\ETOL$ systems except there are no productions in any table from letters of $\Sigma$, $V$ and $\Sigma$ are disjoint, and the derivation relation rewrites only the nonterminals and keeps terminals. With this derivation, a string of terminals cannot be rewritten. We can similarly define the concepts of index $k$ (which uses the number of nonterminals rather than the number of active symbols), and ambiguity. The name {\em reduced} is due to the derivation trees being ``reduced'' as terminals do not rewrite and instead the derivation trees are cut off.

Given any $\ETOL$ system $G$ in active normal form, a reduced $\ETOL$ system, $G'$ can be constructed that simply omits all terminal productions, and in this case, $L(G) = L(G')$, and $G'$ has the same index. However, the concept of ambiguity is arguably simpler with this definition. Certainly, such a reduced $\ETOL$ system does not have to have infinite degree of ambiguity. 

\begin{example} Consider the reduced $\ETOL$ system
$G = (V,{\cal P},S, \Sigma)$ where $\Sigma = \{a,b,\#\}$, ${\cal P} = \{P_S, P_a, P_b, P_f\}$, 
$P_S  = \{S \rightarrow X \# X\}, P_a = \{X \rightarrow aX\}, P_b = \{X \rightarrow bX\}, P_f = \{X \rightarrow \lambda\}$. Then, $L(G) = \{w \# w \mid w \in \{a,b\}^*\}$. Furthermore, it is evident that $G$ is unambiguous, which is certainly more natural than the infinite ambiguity for the corresponding $\ETOL$ system in active normal form. 
\label{normalterminals}
\end{example}

Furthermore, the following is true.
\begin{proposition}
Given an $\ETOL$ system $G = (V, {\cal P},S,\Sigma)$, there exists a reduced $\ETOL$ system $G' = (V', {\cal P}',S', \Sigma)$ with $L(G) = L(G')$ and $\amb(G) \geq \amb(G')$. Moreover, the index of $G$ is equal to the index of $G'$.
\label{makereduced}
\end{proposition}
\begin{proof}
Let $F$ be a new ``dead'' nonterminal, and for all terminals $A \in A_G$, make a new primed nonterminal $A'$. In all productions of every table of ${\cal P}$, replace $A \in A_G \cap \Sigma$ with $A'$ in both left and right hand sides of productions; keep non-active terminals as is on right hand sides, but remove any production rewriting non-active terminals (which must get rewritten to themselves in $G$). Furthermore, make a new production table $P_{\$}$ that maps all primed symbols $A'$ (representing active terminals) to their unprimed variant $A$, and changes all other nonterminals in $V - \Sigma$ to $F$ (which cannot be used in any derivation tree of a word in $L(G)$), and rewrites $F$ to itself. Then, $G'$ simulates $G$ directly, but without rewriting non-active terminals, and at any point where the sentential form contains only primed letters of letters in $A_G \cap \Sigma$ and unprimed terminals, it can change the active symbols to terminals. As there is at most one derivation tree of $G'$ corresponding to each of $G$ (see e.g.\ Figure \ref{etoltoreduced}), the degree of ambiguity is not increased in $G'$. 
\begin{figure}[htbp!]
\begin{center}
\begin{subfigure}{.4\textwidth}
\centering
\includegraphics[width=.8\linewidth]{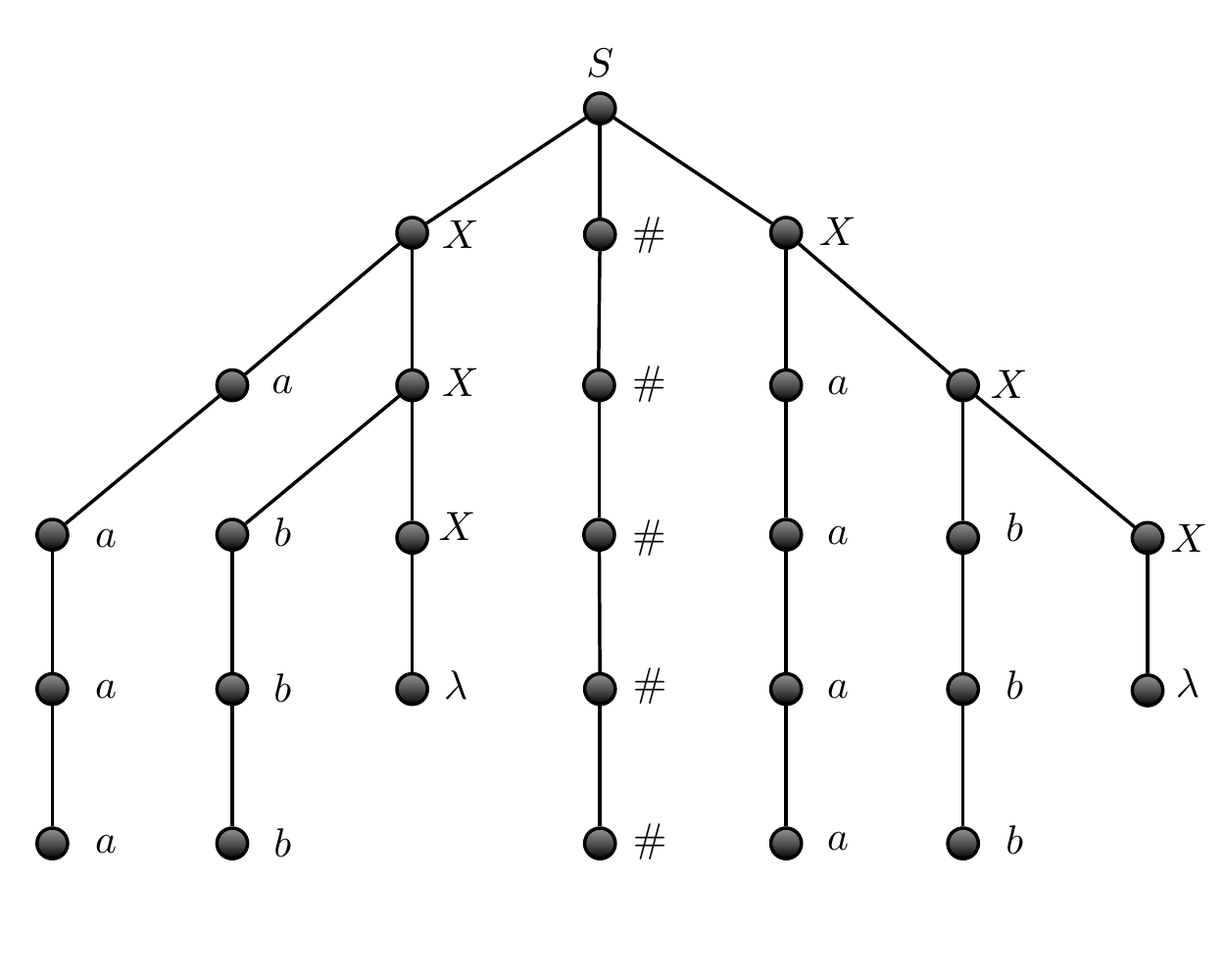}
\caption{derivation tree of $G$}
\label{fig:sfig1}
\end{subfigure}
\begin{subfigure}{.4\textwidth}
\centering
\includegraphics[width=.8\linewidth]{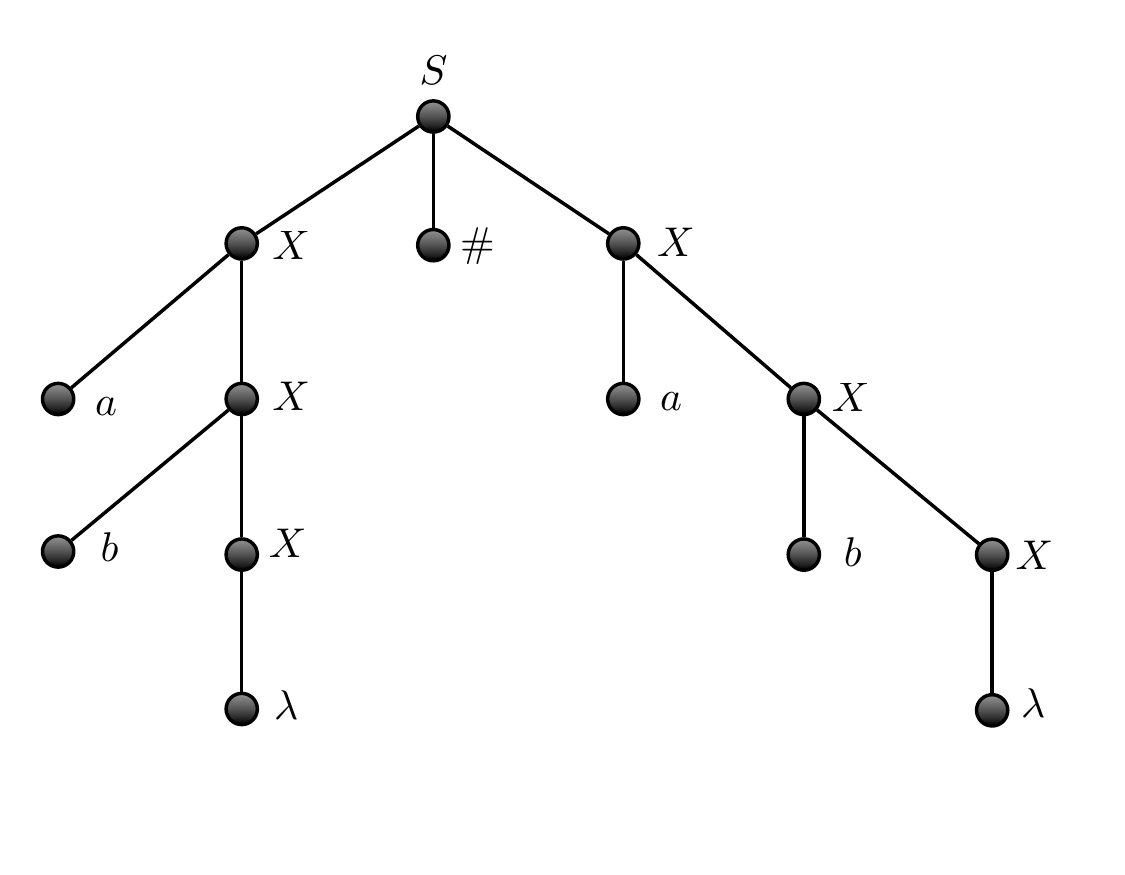}
\caption{derivation tree of $G'$}
\label{fig:sfig2}
\end{subfigure}
\caption{Starting with an $\ETOL$-system generating $\{w \# w \mid w \in \{a,b\}^*\}$ (similar to Example \ref{normalterminals} but with additional productions), from a derivation in image \ref{fig:sfig1}, another in $G'$ is constructed by Proposition \ref{makereduced} in image \ref{fig:sfig2} with the same yield.}
\label{etoltoreduced}
\end{center}
\end{figure}
Furthermore, as the index is defined involving the number of active symbols, the construction changes each active symbol to a nonterminal, which does not affect the index.
\qed	
\end{proof}

The converse also works, which demonstrates that examining reduced $\ETOL$ systems is not necessary in general.
\begin{proposition}
Let  $G = (V,{\cal P},S,\Sigma)$ be a reduced $\ETOL$ system. Then 
	there exists an $\ETOL$ system $\overline{G} = (\overline{V},{\overline{\cal P}}, \overline{S}, \Sigma)$ with $L(G) = L(\overline{G})$, and $\amb(G) = \amb(\overline{G})$, and the index of $G$ is less than or equal to the index of $\overline{G}$.
\label{makenotereduced}
\end{proposition}
\begin{proof}
Let ${\cal P} = \{P_1, \ldots, P_n\}$.
	Define $\overline{V} = V \cup \{\overline{A} \mid A \in V \cup \Sigma \cup \{\lambda\}\} \cup \{A_1, A_2 \mid A \in \Sigma \cup \{\lambda\}\} \cup \{F\}$ (where $\overline{A}, A_1,A_2$ are new symbols, as are $\overline{\lambda},\lambda_1,\lambda_2$),  
	$\overline{\cal P} = \{
	\overline{P_1}, \ldots, \overline{P_n}, P_{ \$} \}$. Let $h_1$ be a function from $(V \cup \Sigma)^*$ to $\overline{V}^*$ that, akin to a morphism, maps each letter of $V$ to itself, and each $a \in \Sigma$ to $a_1$, but also maps $\lambda$ to $\lambda_1$ (which makes it not a morphism). Similarly, for a word $w \in \Sigma^*$, let $\overline{w}$ be the word obtained by adding a bar to each letter of $w$, but on $\lambda$, changes it to $\overline{\lambda}$. The construction of productions is as follows:
	\begin{itemize}
		\item For all $A \rightarrow \alpha \in P_i$, to $\overline{P_i}$, add: $A \rightarrow h_1(\alpha)$. Furthermore, if $\alpha \notin \Sigma^*$ (and so there is some letter in $V$ in $\alpha$), add all possible productions of the form $\overline{A} \rightarrow \alpha'$, where $\alpha'$ is obtained from $h_1(\alpha)$ by replacing one or more occurrences of any letter $B \in V$ with $\overline{B}$. If $\alpha \in \Sigma^*$, add $\overline{A} \rightarrow \overline{\alpha}$.
		\item In all $\overline{P_i}$, for all $a \in \Sigma\cup \{\lambda\}$, add $a_1 \rightarrow a_2, a_2 \rightarrow a_2, a \rightarrow F, F \rightarrow F$.
		\item In $P_{\$}$, add $F \rightarrow F$; for all $a \in \Sigma \cup \{\lambda\}$, add $\overline{a} \rightarrow a, a_1 \rightarrow F, a_2 \rightarrow a$; for all $a \in \Sigma$, add $a \rightarrow F$; for all $A \in V$, add $A \rightarrow F, \overline{A} \rightarrow F$.
	\end{itemize}
	
	Essentially, $\overline{G}$ simulates $G$, and in every derivation, $\overline{G}$ guesses one or more paths (later verified to be every path with maximum height in the derivation trees in $G$), and uses bars on each letter derived on those paths. Only $P_{\$}$ can be used to derive terminals, and therefore it must be used last. Furthermore, if it is used once and does not produce all terminals, then at least one $F$ must be produced, and it can never change from $F$ and therefore never produces a word in the language. Therefore, it can only be used once at the last step of the derivation, only producing letters in $\Sigma$ in any successful derivation. Furthermore, for all non-barred nonterminals produced, any terminals produced use a $1$ as a subscript, which must then immediately change to $2$ as subscript at the next height, and then they remain the same until the last step of the derivation where they change into the appropriate terminal.  As $P_{\$}$ changes any letter with $1$ as subscript to $F$, any production of $G$ simulated that is not barred that produces a terminal or the empty word in $G$, must be simulated before the last simulated step of the derivation in $\overline{G}$. For this reason, the barred terminals must be exactly those produced at the last step of the derivation. This has the effect of pushing all terminals produced to the last height of every derivation tree, but the heights of the derivation trees are always one more (due to the application of $P_{\$}$) than the corresponding heights in $G$. 	
	An example of the conversion is shown in Figure \ref{etolremovereduced}.
\begin{figure}[htbp!]
\begin{center}
\begin{subfigure}{.4\textwidth}
\centering
\includegraphics[width=.8\linewidth]{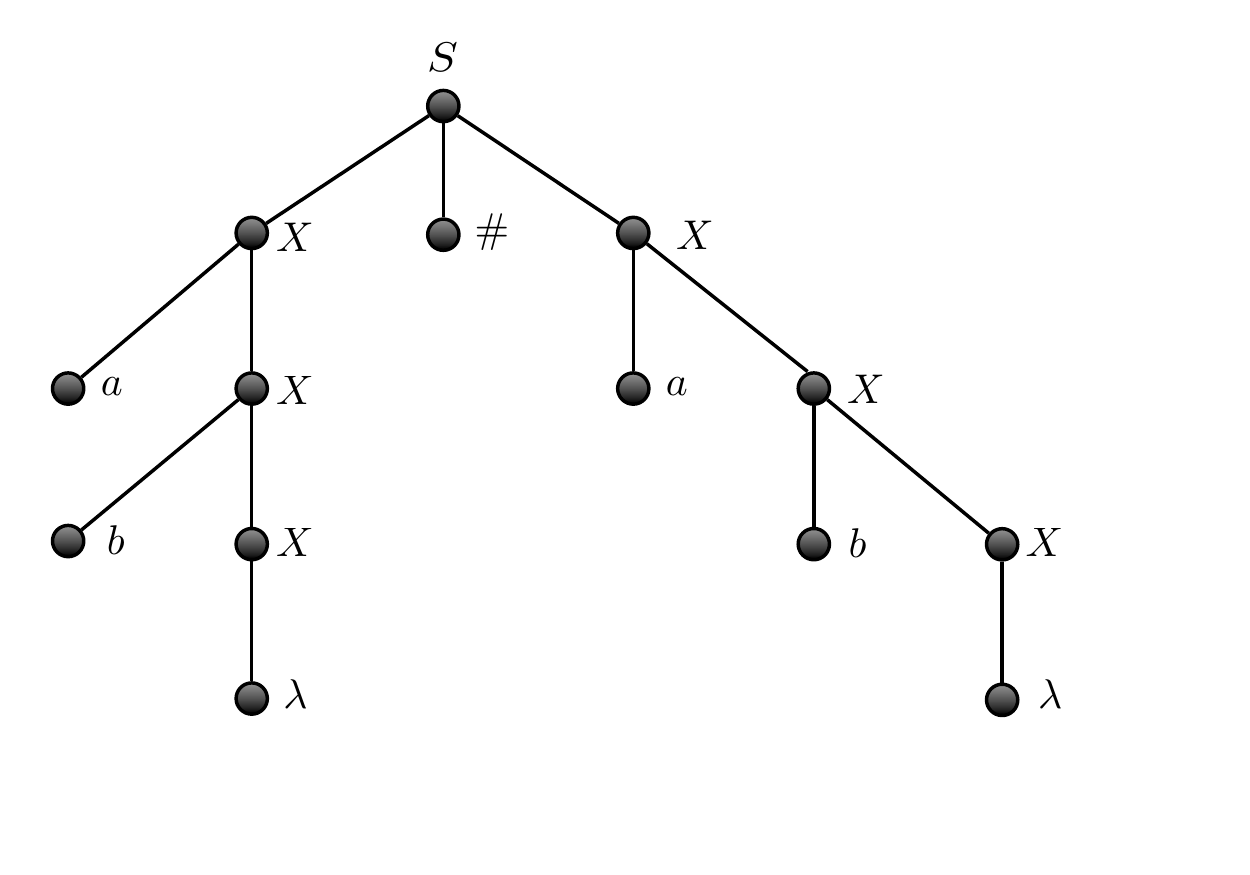}
\caption{derivation tree of $G$}
\label{fig:sfig3}
\end{subfigure}
\begin{subfigure}{.4\textwidth}
\centering
\includegraphics[width=.8\linewidth]{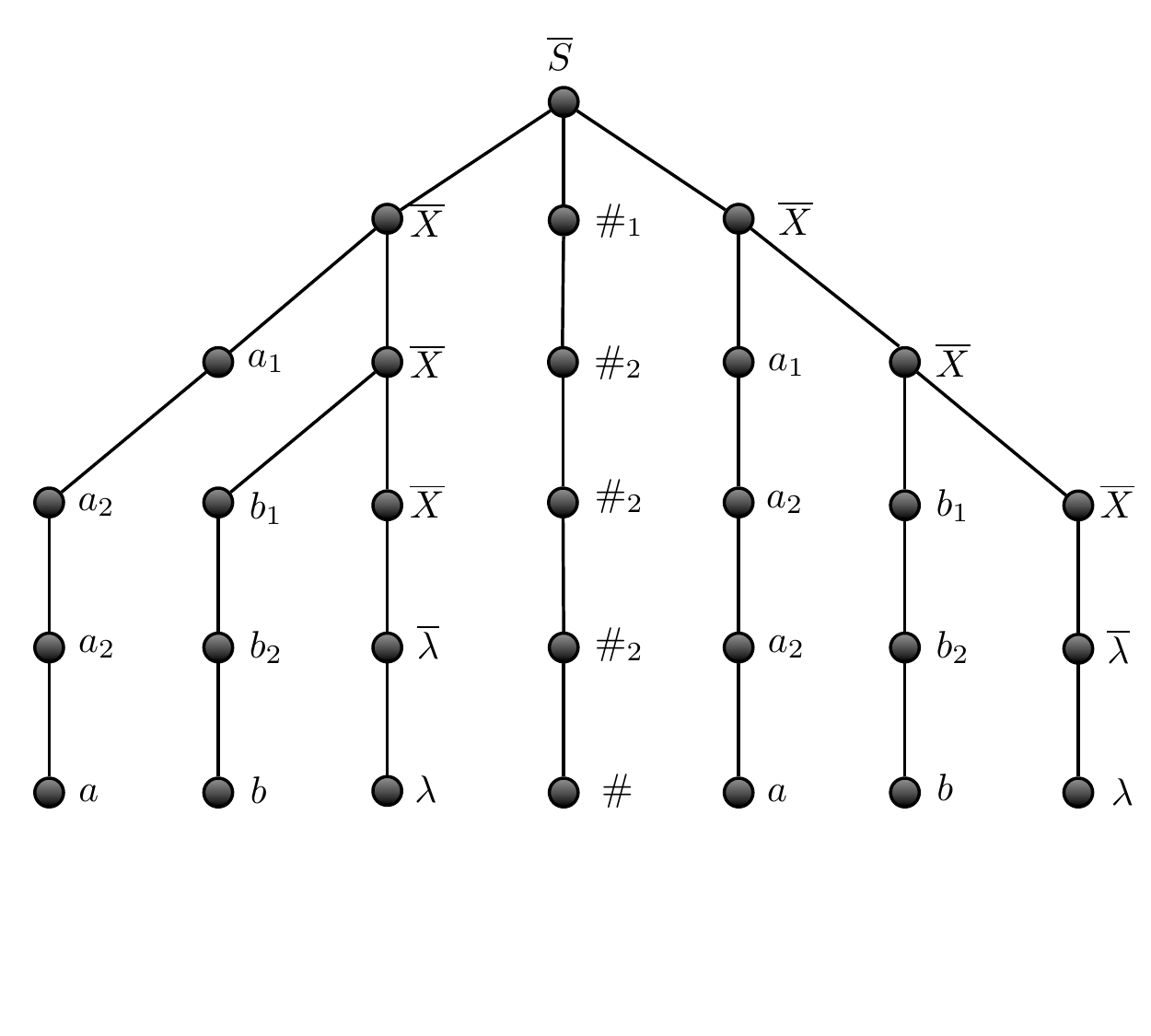}
\caption{derivation tree of $G'$}
\label{fig:sfig4}
\end{subfigure}
\caption{Starting with the reduced $\ETOL$-system generating $\{w \# w \mid w \in \{a,b\}^*\}$ in Example \ref{normalterminals}, from a derivation in image \ref{fig:sfig3}, a non-reduced $\ETOL$-system in $G'$ is constructed by Proposition \ref{makenotereduced} in image \ref{fig:sfig4} with the same yield.}
\label{etolremovereduced}
\end{center}
\end{figure}

The degree of ambiguity is identical here as there is a bijective correspondence between derivation trees. This is due to the guessed longest paths, which enforces that there is only one tree in $G'$ corresponding to each tree in $G'$ (of one height longer). Had this not been present, each symbol of the form $a_2$ could keep rewriting to itself indefinitely producing an infinite number of trees corresponding to each in $G$.
	In addition, the index cannot decrease with this procedure.
\qed
\end{proof}

\begin{proposition} \label{ambiguoussameforreduced}
A language $L \in \LL(\ETOL)$ can be generated by an $r$-ambiguous $\ETOL$ system if and only if it can be generated by an $r$-ambiguous reduced $\ETOL$ system.
\end{proposition}
\begin{proof}
Consider $L$ generated by an $r$-ambiguous $\ETOL$ system $G$, where $r$ is minimal. By Proposition \ref{makereduced}, there
exists a $r'$-ambiguous reduced $\ETOL$ system $G'$ generating $L$ with $r' \leq r$. Suppose $r' < r$. By Proposition
\ref{makenotereduced}, there exists an $r'$-ambiguous $\ETOL$ system $G''$ generating $L$, contradicting minimality. Thus, $r = r'$. The other direction follows directly from Proposition \ref{makenotereduced}.
\qed
\end{proof}

From these results, it is enough to use reduced $\ETOL$ to study ambiguity, which simplifies the discussion and makes the comparison to other grammar systems simpler.
However, notice in the construction of Proposition \ref{makenotereduced}, even if $G$ is finite-index, $\overline{G}$ is not necessarily finite-index. For example,  as each of the symbols $a_1, a_2$ etc.\ created from terminals are active symbols, these symbols are used in place of terminals. However, the terminal symbols $a \in \Sigma$ do not count towards the index in $G$, but the symbols $a_1,a_2$ are active symbols in $\overline{G}$ as they can get rewritten to $a$ at the last step of the derivation, and therefore count towards the index of $\overline{G}$. In fact, the following is true.
\begin{proposition} Let $L$ be any finite-index $\ETOL$ language with two letters $a,b$ such that $\{ |w|_a \mid w \in L \}$
and $\{ |w|_b \mid w \in L\}$ are infinite.
Every finite-index $\ETOL$ system $G$ that generates $L$ is ambiguous of degree $\infty$.
However, there exist some such languages $L$ that are generated by a finite-index finitely ambiguous reduced
$\ETOL$ system.
\label{usereduced}
\end{proposition}
\begin{proof}
Assume otherwise, and there is a finitely ambiguous $\ETOL$ system $G$ of index $k$. 
Consider a successful derivation $S \Rightarrow^* x \in \Sigma^*$ where $|x|_a>k$. Since there are more than $k$ $a$'s in $x$, $a$ must not be an active symbol. A similar derivation can be used to show that $b$ must not be an active symbol. Thus, $a\rightarrow a$ and $b \rightarrow b$ must be in every table. By Remark \ref{neverambiguous}, $G$ cannot be finitely-ambiguous.

This is the case say for $L$ in Example \ref{normalterminals}  (without the $\#$ symbol) --- there is no finitely ambiguous finite-index $\ETOL$ system $G$ that generates $L$, but there is a reduced $\ETOL$ system which is unambiguous and of index $2$.
	\qed
\end{proof}

Thus, given any $\ETOL$ language $L$, there is an $\ETOL$ system $G$ accepting $L$ with a finite degree of ambiguity if and only if there is a reduced $\ETOL$ system $G'$ accepting $L$ with a finite degree of ambiguity. However, the index of $G'$ is less than or equal to $G$, and sometimes $G'$ is finite-index but $G$ is not. 
 Furthermore, there are some finite-index $\ETOL$ languages $L$ where there is a finite-index unambiguous reduced $\ETOL$ system generating $L$, but there is no finite-index finitely-ambiguous $\ETOL$ system generating $L$. 
Hence, in terms of finite-index $\ETOL$, measuring ambiguity with reduced $\ETOL$ is a strictly stronger method. Therefore, in the sequel, we will adopt this method.

Next we will show that this notion of $r$-ambiguity for reduced $\ETOL$ is equivalent to the notion for matrix grammars.
\begin{lemma}
There exists an algorithm which, given an arbitrary matrix grammar $G$ of index $k$ that is $r$-ambiguous, produces a reduced $\ETOL$ system $G'$ of uncontrolled index $k$ with ambiguity $r$.
\end{lemma}
\begin{proof}
This is similar to Lemma 5 from \cite{RozenbergFiniteIndexGrammars}. Let $G = (N, \Sigma, M, S)$ be one such matrix grammar. Define a reduced $\ETOL$ system $G' = (V,{\cal P}, S, \Sigma)$ with
$V = \{[x,i] \mid x \in N^{\leq k}, 1 \leq i \leq |x|\} \cup \{F\}$ where $F$ is a ``dead'' nonterminal. For $x \in N^{\leq k}$ and $m \in M$, let $Der(x,m)$ be the set of all $\{(w_1, \ldots, w_{|x|}) \mid x \Rightarrow_m w_1 \cdots w_{|x|}$, and the $i$th letter of $x$ derives $w_i\}$. 
For each $x \in V^{\leq k}, m \in M, y = (w_1, \ldots , w_{|x|}) \in Der(x,m)$,
make a table $T_{x,m,y}$; 
and for each $i$, $1 \leq i \leq |x|$ letting $t = | \pi_N(w_1 \cdots w_{i-1})|$ and $w_i = y_0 A_1 \cdots A_n y_n, y_j \in \Sigma^*, A_j \in N, n \geq 0$, the table contains
$$[x,i] \rightarrow y_0 [A_1, t+1] y_1 \cdots [A_n,t+n] y_n,$$
and create $X \rightarrow F$ for all $X \in V \cup \{F\}$ not of this form.

Assume $G$ is $r$-ambiguous. Then given each $w \in \Sigma^*$, there are at most $r$ derivations.
Given each derivation $x_0 \Rightarrow_{m_1} x_1  \cdots \Rightarrow_{m_l} x_l$, then there is a derivation of $G'$ associated with this, but it's also a bijection since it can be inverted.
\qed
\end{proof}

\begin{lemma}
There exists an algorithm which, given a reduced $\ETOL$ $G$ of index $k$ that is $r$-ambiguous, produces an equivalent reduced
$\EDTOL$ $G'$ of index $k$ that is $r$-ambiguous. 
\end{lemma}
\begin{proof}
This is essentially the same as Lemma 2 from \cite{RozenbergFiniteIndexETOL}. That is, for every $x \in V^{\leq k}$, it labels the nonterminals with uniquely occurring subscripts from $\{1, \ldots, k\}$. And for every $P$ and every combination of ways of rewriting $x$, it makes a new production table. Clearly, this maps derivations in a bijective fashion, preserving ambiguity.
\qed
\end{proof}

\begin{lemma}
There exists an algorithm that, given an arbitrary reduced $\ETOL$ of index $k$ that is $r$-ambiguous, produces an equivalent matrix grammar of uncontrolled index $k$ that is $r$-ambiguous.
\end{lemma}
\begin{proof}
The construction from \cite{RozenbergFiniteIndexGrammars} is bijective.
\qed
\end{proof}

These can be summarized as follows:
\begin{proposition}\label{equiv-fin-amb-ET0L-MatrGram}
The $k$-index $r$-ambiguous matrix languages are equal to the $k$-index $r$-ambiguous reduced $\ETOL$ languages.
\end{proposition}

\section{On the Characteristic Series of Finite-Index Matrix Grammars}\label{sec:Char-Series-Matrix-Grammar}

The goal of this section is to prove that if a language $L$ is generated by an unambiguous context-free matrix grammar of finite index, then the characteristic series in commutative variables of $L$ is rational and $L$ is counting regular.


The following lemma, concerning grammars of finite index, will be useful for our purposes.

\begin{lemma}
	\label{FIML-useful-gr}     
	For each matrix grammar $G$ of index $k$, one can construct a matrix grammar $G'$ of index $k$ such that $L(G) = L(G')$, and:
	\begin{enumerate}
		\item
		all sentential forms derivable in $G'$ contain distinct nonterminal occurrences;
		\item
		if one applies some matrix $(A_1 \rightarrow w_1, \ldots, A_i \rightarrow w_i)$ to a string $\nu$, then all $A_1, \ldots, A_i$ occur in $\nu$
		and each rule $A_j \rightarrow w_j$, replaces an occurrence of $A_j$ in $\nu$.
	\end{enumerate}
    Moreover, if the grammar $G$ is unambiguous, then $G'$ is unambiguous, too.
\end{lemma}

A proof of the previous lemma can be found in~\cite[Lemma 3.1.4]{DassowPaun}
without the final statement concerning unambiguity.
However, by inspecting the proof, one is easily convinced that the construction of $G'$ preserves unambiguity.

We will say that a finite-index matrix grammar is in \emph{normal form} if it satisfies Conditions 1 and 2 of Lemma~\ref{FIML-useful-gr}.
We notice that a grammar of finite index in normal form $G$ satisfies the following properties:
\begin{itemize}
	\item[P1.] 
	\emph{For all $\alpha \in M^*$, there is at most one $x\in S(G)$ such that $S\derivestep[\alpha]x$.}
	\item[P2.]
	\emph{The set $\pi_N(S(G))$ is finite.}
\end{itemize}

The elements of $M^*$ may be viewed as words on the alphabet $M$.
Thus, subsets of $M^*$ are formal languages.
In particular, we are interested in the language
\[
D(G)=\{\alpha \in M^*\mid S\derivestep[\alpha]w,\ w\in L(G)\},
\]
which, in a certain sense, represents the derivations of the grammar $G$.
This language is known as the {\em Szilard language} of the matrix grammar. The following was shown by
P\u{a}un:
\begin{proposition} \cite{PaunMatrix2} \label{prop:Proposition15} 
	Let $G$ be a finite-index matrix grammar.
	Then $D(G)$ is a regular language.
\end{proposition}
We assume the reader to be familiar with the theory of rational series on a monoid and of recognizable subsets of a monoid (see, e.g.,~\cite{SalomaaSoittola,Eilenberg}).
Here, we limit ourselves to fix some notation and recall some results useful for our purposes.
Moreover, we will consider only  series with coefficients in the complete semiring $\Nc=\mathbb N\cup\{+\infty\}.$

A formal $\Nc$-\emph{series} on a monoid $M$  (or $\Nc$-\emph{subset} of $M$) is any map $\mathcal S\colon M\to\Nc$.
The image of any $w\in M$ by $\mathcal S$ is usually denoted by $(\mathcal S, w)$ and is called the \emph{coefficient} (or \emph{multiplicity}) of $w$ in $\mathcal S$.

Let $M$ and $M'$ be two monoids, $\theta\colon M\to M'$ a monoid morphism and $\mathcal S$ a rational series on $M$.
Then the formal series $\theta(\mathcal S)$ defined on $M'$ by
\[
(\theta(\mathcal S),v)=\sum_{w\in\theta^{-1}(v)}(\mathcal S,w),\quad
v\in M',
\]
is rational.

Let $\mathcal S$ be a rational series on a monoid $M$ and $R$ a recognizable subset of $M$.
Then the series $\mathcal S\cap R$ defined on $M$ by
\[
(\mathcal S\cap R,w)=\begin{cases}
(\mathcal S,w),&\mbox{if }w\in R,\\
0,&\mbox{if }w\in M\setminus R,
\end{cases}
\]
is rational
(see~\cite[Proposition VII, 5.3]{Eilenberg})

Let $L$ be a subset of $\Sigma^*$.
The \emph{characteristic series} of $L$ in \emph{non-commutative variables} is the series $\underline L$ with coefficients
\[
(\underline L,w)=\begin{cases}
1,&\mbox{if }w\in L,\\
0,&\mbox{otherwise,}
\end{cases}
\]
where $w\in\Sigma^*$. As a consequence of Kleene's Theorem, the series $\underline L$ is rational if and only if the language $L$ is regular.

We denote by $\sim$ the commutative equivalence of $\Sigma^*$, that is, the congruence of $\Sigma^*$ generated by
\[
ab\sim ba \quad\mbox{for all }a,b\in \Sigma.
\]
The \emph{characteristic series} of $L$ in \emph{commutative variables} is the series $\uuline L=c(\underline L)$, where
$$c\colon\Sigma^*\to\Sigma^\oplus=\Sigma^*/\sim$$ is the natural projection of $\Sigma^*$ into the free Abelian monoid generated by $\Sigma$.
In other words, the characteristic series of $L$ over the alphabet $\Sigma=\{a_1, \ldots,a_n\}$ is the formal series
$$
\sum \alpha {a_1}^{i_1}\cdots {a_n}^{i_n},
$$ 
where $\alpha$ is the number of words of $L$ whose Parikh image is equal to the vector $(i_1,\dots, i_n)$.



Let $G$ be a grammar.
 The \emph{characteristic series} of $G$ in non-commutative variables is the formal series $\underline G$ on $\Sigma^*$ whose coefficients $(\underline G,w)$ count the number (possibly infinite) of distinct derivations of $w$ in $G$, $w\in\Sigma^*$.
The series $$\uuline G=c(\underline G)$$ is the \emph{characteristic series} of $G$ in commutative variables.

 
We will show that the characteristic series of a finite-index matrix grammar in normal form is a rational series.
In order to achieve this result, we introduce the morphism $\theta\colon M^*\to \Sigma^*$ defined as follows.
For any matrix rule $m=(X_1\rightarrow\gamma_1,\ldots,X_h\rightarrow\gamma_h)$, one has $\theta(m)=\pi_{\Sigma}(\gamma_1\cdots\gamma_h)$.
The following lemma states a useful property of the morphism $\theta$.
\begin{lemma}
	\label{lem:phi}
	If one has $S\derivestep[\alpha]v$, $v\in L(G)$, $\alpha \in D(G)$, then $v\sim\theta(\alpha)$.
\end{lemma}
\begin{proof}
	If $S\derivestep[\alpha ]v$, then there are $h>0$,
	$m_1,\ldots,m_h\in M$, $x_0,x_1,\ldots,x_h\in S(G)$, such that
	\[
	\alpha =m_1\cdots m_h,\quad
	S=x_0\derivestep[m_1] x_1\derivestep[m_2]\cdots\derivestep[m_h] x_h=v.
	\]
	From the definition of $\theta$ one easily derives that
	\[
		\pi_{\Sigma}(x_{i-1})\theta(m_i)\sim\pi_{\Sigma}(x_i),\quad
		1\leq i\leq h.
	\]
	It follows that $\theta(\alpha)=\pi_{\Sigma}(x_{0})\theta(m_1m_2\cdots m_h)\sim\pi_{\Sigma}(x_{h})=v$.\qed
\end{proof}

The following proposition extends a result of \cite{BaronKuich} to matrix grammars.

%
%

\begin{proposition}
	\label{prop:razionale}
	Let $G$ be a finite-index matrix grammar in normal form.
	The series $\uuline G$ is rational.
\end{proposition}
\begin{proof}
Consider the series (in commutative variables) $\mathcal D=c\big(\theta\big(\underline{ D(G)}\big)\big)$.
By  the regularity of $D(G)$ 
and the fact that morphisms preserve rational series, $\mathcal D$ is rational.
Thus, to prove our statement, it is sufficient to verify that $\mathcal D=\uuline G$.
 
	Using Lemma~\ref{lem:phi}, one can easily check that, for all $w\in {\Sigma}^*$,
	\begin{equation}\label{eq:primo}
		\{\alpha \in D(G)\mid \theta(\alpha)\sim w\}=	
		\{\alpha \in D(G)\mid S\derivestep[\alpha]v,\ v\sim w,\ v\in L(G)\}.
	\end{equation}

Let $w$ be any word of $\Sigma^*$ and $\widetilde{w}=c(w)$.
Notice that the coefficient of $\widetilde w$ in $\mathcal D$ is equal to the cardinality of the left hand side of (\ref{eq:primo}) while, in view of Condition P1, the coefficient of $\widetilde w$ in $\uuline G$ is equal to the cardinality of the right hand side of (\ref{eq:primo}).
%
	The conclusion follows.\qed
\end{proof}

It is easily seen that a matrix grammar $G$ is unambiguous if and only if
\[
\uuline{\vphantom)G}=\uuline{L(G)}.
\]

As a straightforward consequence of Lemma~\ref{FIML-useful-gr} and Proposition~\ref{prop:razionale} we obtain our main results.
\begin{proposition}
\label{umambiguous-matrix-rational}
	Let $L$ be the language generated by an unambiguous finite-index matrix grammar.
	The series $\uuline L$ is rational.
\end{proposition}

We recall that the \emph{generating series} of a language $L$ is the formal series $\sum_{w\in L}x^{|w|}$.
Since the generating series of a language is the morphic image of the characteristic series, one has the following.

\begin{corollary}
\label{matrixrational}
	Let $L$ be the language generated by an unambiguous finite-index matrix grammar.
	The generating series of $L$ is rational. Therefore, $L$ is counting regular.
\end{corollary}

In view of Proposition \ref{equiv-fin-amb-ET0L-MatrGram}, the latter results now implies the following proposition for  reduced finite-index $\ETOL$ 
systems.
\begin{proposition}
	Let $L$ be an unambiguous  reduced  finite-index $\ETOL$ language.   		
	The series $\uuline L$ is rational.
	In particular, the generating series of $L$  is rational, and $L$ is counting regular.
\end{proposition}

   Let $n\geq 2$ and let $\Sigma_n = \{a_1, \ldots, a_n\}$ be an alphabet of $n$ symbols. Let ${\cal L}_n$ be the language over $\Sigma_n$ defined as:
   $${\cal L}_n = \{x\in \Sigma_n \mid |u|_{a_1}=|u|_{a_2}= \cdots = |u|_{a_n}\}.$$
Observing that ${\cal L}_2$ is the language of  $2$-sided Dyck words whose generating series  is not rational, one gets.
\begin{corollary}
\label{NotMatrix}
	 ${\cal L}_2$ cannot be generated by an unambiguous finite-index matrix grammar.
Similarly, ${\cal L}_2$ cannot be generated by an unambiguous  reduced  finite-index $\ETOL$ system.
\end{corollary}

\begin{proposition}
The unambiguous $\NCM$ languages and the unambiguous $\M\fin$ languages (unambiguous reduced $\ETOL\fin$ languages) are incomparable.

\end{proposition}
\begin{proof}
Consider $L = \{ x \# x \mid x \in \{a,b\}^+ \}$. It is easy to build a unambiguous finite-index matrix grammar generating $L$. Indeed,
consider the following matrices: $\big[  S \rightarrow  A \# B\big],
\big[A \rightarrow a A,  B \rightarrow aB\big],
\big[A \rightarrow bA,  B \rightarrow bB\big],
\big[A \rightarrow a,  B \rightarrow a\big],
\big[A \rightarrow b,  B \rightarrow b\big]$.
But it follows from Lemma \ref{strictness} that $L$ is not in $\LL(\NCM)$.

Consider $\LL_2 = \{ x \mid |x|_{a_1} = |x|_{a_2}\}$. It is easy to build a $\DCM$ (hence an unambiguous $\NCM$) accepting $L$. But,
$L$ is not an unambiguous finite-index matrix language by Corollary \ref{NotMatrix}.
\qed \end{proof}

We observe that for every $n\geq 2$, the characteristic series $\uuline{\mathcal L_k}$ is not rational.
Indeed,  let $\Sigma_n^\oplus=\Sigma_n^*/\sim$ be the free Abelian monoid generated by $\Sigma_n$, $n\geq 3$.
The submonoid $\Sigma_2^\oplus$ is a recognizable subset of the monoid $\Sigma_n^\oplus$.
Moreover, as one easily verifies,
$\uuline{\mathcal L_2}=\uuline{\mathcal L_n}\cap \Sigma_2^\oplus$.
We derive that the series $\uuline{\mathcal L_n}$ cannot be rational since, otherwise, also 
$\uuline{\mathcal L_2}$ should be rational and this is not the case.
Thus, the latter corollary can be extended to all languages ${\cal L}_n$, $n\geq 2$.

%
%
%
%

\section{Inherent Ambiguity}
\label{sec:inherent}

The results of the previous section connect directly to the important notion of inherent ambiguity.
Given a class of grammars ${\cal G}$, a language $L$
is said to be {\em inherently ${\cal G}$-ambiguous} if
$L \in \LL({\cal G})$ and every $G \in {\cal G}$ such that $L = L(G)$ is ambiguous. Similarly, given a class of
machines ${\cal M}$, a language $L$ is {\em inherently ${\cal M}$-ambiguous} if 
$L \in \LL({\cal M})$ and every $M \in {\cal M}$ such that $L = L(M)$ is ambiguous.
For simplicity, we say a $\ETOL\fin$ language is inherently {\em $\ETOL\fin$-ambiguous} if every 
reduced $\ETOL\fin$ system $G$ generating $L$ is ambiguous.
We use reduced $\ETOL\fin$ for this definition, as it is a stronger and more useful definition than $\ETOL\fin$
generally due to the results in Proposition \ref{usereduced}. For $\ETOL$ generally, the reduced
concept is equally as strong by Proposition \ref{ambiguoussameforreduced}.
This concept, to our knowledge has never been defined or studied for finite-index $\ETOL$, finite-index matrix languages, or $\ETOL$ generally, but has been studied for $\EOL$ systems. In \cite{InherentAmbiguityE0L}, it was shown that
there exists an inherently $\EOL$-ambiguous language.

Two types of problems are of interest with respect to inherent ambiguity. 
\begin{problem}
For a given class of grammars ${\cal G}$ (or machines ${\cal M}$), does there exist any inherently ${\cal G}$-ambiguous  
(inherently ${\cal M}$-ambiguous resp.) languages? 
\end{problem}
For example, for the class of
context-free grammars $\CFG$, it is well-known that there are inherently $\CFG$-ambiguous languages \cite{Parikh}.

Next, it was conjectured by Chomsky that there are two sub-classes of context-free
grammars ${\cal A}$ and ${\cal B}$, with ${\cal A} \subsetneq {\cal B}$, such that there are languages
that are inherently ${\cal A}$-ambiguous, but not inherently ${\cal B}$-ambiguous \cite{Chomsky}. This has been confirmed for
some classes. There exist such languages when ${\cal A}$ is the class of linear context-free grammars with just one
nonterminal and ${\cal B}$ is the class of  linear context-free grammars \cite{Gross}.
It is also true when ${\cal A}$ is the linear context-free grammars and ${\cal B}$ is the context-free grammars \cite{Blattner}.
More generally, for all grammar forms \cite{GrammarForms} that describe proper subsets ${\cal A}$ of the
context-free grammars, there is some language that is inherently ${\cal A}$-ambiguous but not inherently $\CFG$-ambiguous. Such a form includes the $k$-linear grammars \cite{GrammarForms}, which generate the finite union of products
of $k$ linear context-free languages. These can describe a strict subset of the finite-index context-free grammars (also called
derivation-bounded context-free grammars) \cite{FiniteTurn}. 
Hence, the second problem is:
\begin{problem}
\label{twoclasses}
For two classes of grammars (or machines) ${\cal A}$ and ${\cal B}$, with ${\cal A} \subsetneq {\cal B}$,
does there exist any languages that are inherently ${\cal A}$-ambiguous, but not inherently ${\cal B}$-ambiguous?
\end{problem}

It is evident from Proposition \ref{equiv-fin-amb-ET0L-MatrGram} that a language is inherently $\M\fin$-ambiguous if and only if it is inherently $\ETOL\fin$-ambiguous.
Furthermore, there exist $\M\fin$ languages whose generating series are algebraic not rational and, even, transcendental, as proven by Flajolet in \cite{fla}. 
   For example, the generating series of the inherently  ambiguous linear context-free language  of exponential growth 
   $$S = \{a^nbv_1a^nv_2 \mid n\geq 1, v_1, v_2\in \{a,b\}^*\}$$ is transcendental ({\em cf.}  \cite{fla}, Theorem 3). 
From this, we are able to determine the following with an easy application of
Corollary \ref{matrixrational}:
\begin{proposition}
There exist inherently $\M\fin$-ambiguous and inherently $\ETOL\fin$-ambiguous languages.
\end{proposition}
\begin{proof} 
Consider the language $S$ above which is an $\M\fin$ and $\ETOL\fin$ language. Assume $S$ is not inherently $\M\fin$-ambiguous.
Thus, there exists an unambiguous $\M\fin$ grammar $G$ generating $S$. The generating series of $S$ is not rational
\cite{fla}. But by Corollary \ref{matrixrational}, the generating  series of $S$ must be rational, a
contradiction. Therefore, $S$ is an inherently $\M\fin$-ambiguous (and inherently $\ETOL\fin$-ambiguous) language.
\qed
\end{proof}
This is the first such language in the literature.

We see next that this also gives a positive solution to Problem \ref{twoclasses}.
\begin{proposition}
There are languages that are inherently $\ETOL\fin$-ambiguous languages but not inherently $\ETOL$-ambiguous.
\label{inherentlyambetol}
\end{proposition}
\begin{proof}
Let $R = \{a^n b (a^{m_1}b) \cdots (a^{m_k}b) a^n v \mid n\geq 1, k \geq 0, m_i <n \mbox{~for~} 1 \leq i \leq k, v\in \{a,b\}^*\}$. First we will prove $S= R$.

Let $w \in S$. Then $w = a^n b v_1 a^n v_2, n \geq 1, v_1, v_2 \in \{a,b\}^*$. In
$v_1 a^n$, there is some first time that $a^n$ occurs, say starting at position $i \geq 1$. Let
$v_1 a^n = x_1 a^n x_2$ where $|x_1| = i-1$, and $x_1 , x_2 \in \{a,b\}^*$.
By the minimality of $i$, $x_1$ cannot end with $a$, and cannot contain $a^n$ as subword. Thus,
$x_1 = (a^{m_1}b) \cdots (a^{m_k} b)$, where $k \geq 0$, and $ 0 \leq m_i < n$ for $1 \leq i \leq k$. Thus,
$w  = a^n b (a^{m_1} b) \cdots (a^{m_k} b) a^n x_2 v_2$, and $w \in R$.

The other direction is clear, hence $S = R$.

We will intuitively describe how to create an unambiguous reduced $\ETOL$ system generating $R$. For
$w = a^n b (a^{m_1}b) \cdots (a^{m_k}b) a^n v$, where $n\geq 1, k \geq 0, m_i <n \mbox{~for~} 1 \leq i \leq k, v\in \{a,b\}^*$,
the unique derivation derives the first $a^n$ in a first path, generates each $a^{m_i}b$ in a path (the number
$k$ and each $m_i$ is uniquely determined given $w$) where it is verified that each $m_i < n$, generates the next $a^n$ in the next path, and has one final path generating $v$.
\qed
\end{proof}

This does not resolve the question of whether there exists any inherently $\ETOL$-ambiguous language, which remains an open problem that has not been studied.

Next, we examine another technique to conclude that languages are inherently ${\cal G}$-ambiguous for certain classes of grammars and machines involving counting languages. 
The following was shown in Theorem 6 and Corollary 7 of \cite{iba-2-journ}.
Let $M$ be a machine from any of the following machine models with a one-way read-only input:
\begin{itemize}
\item unambiguous nondeterministic reversal-bounded Turing machines,
\item unambiguous nondeterministic reversal-bounded pushdown automata,
\item unambiguous nondeterministic reversal-bounded queue automata,
\item unambiguous nondeterministic reversal-bounded $k$-flip pushdown automata.
\end{itemize}
Then $L(M)$ is counting regular.
Here, reversal-bounded pushdown automata have a bound on the number of times it can switch from pushing to popping and vice versa, queue automata have a bound on the number of switches from enqueueing to dequeueing and vice versa, the Turing machines have a single read/write worktape and there is a bound on the number of switches between moving right and left, and vice versa, and $k$-flip are allowed to `flip' their pushdown up to $k$ times but there is a bound on the number of switches between pushing and popping. Hence if a language $L$ can be accepted by one of these models, and it is not counting regular, then it must be inherently ambiguous for that model. 
%
In this case the language $S$ is not counting regular, because 
the generating series of $S$ is not rational \cite{fla}.
Hence, any machine model ${\cal M}$ above (or those that can be unambiguously simulated by the models above) that can accept $S$ must do so ambiguously, and therefore $S$ is inherently ${\cal M}$-ambiguous.
\begin{proposition}
The following classes of machines ${\cal M}$ contain inherently ${\cal M}$-ambiguous languages:
\begin{itemize}
\item nondeterministic reversal-bounded Turing machines,
\item nondeterministic reversal-bounded pushdown automata,
\item nondeterministic reversal-bounded queue automata,
\item nondeterministic reversal-bounded flip pushdown automata,
\item nondeterministic reversal-bounded stack automata,
\item nondeterministic reversal-bounded checking stack automata,
\item $\NCM(1)$,
\item $\NCM(1,1)$.
\end{itemize}
\end{proposition}
Stack automata have a pushdown with the ability to read the contents of the pushdown in read-only mode \cite{CheckingStack}. 
Such a machine is reversal-bounded if there is a bound on the number of switches between moving right (by pushing or moving the read head to the right), or moving left (by popping or moving the read head to the left). Checking
stack automata are further restricted so that once they start reading from the inside of the stack, they no longer can write to it \cite{CheckingStack}. This example is particularly interesting as deterministic checking stack automata can unambiguously accept $S$, thereby providing another solution to the second problem. We provide two other solutions to Problem \ref{twoclasses}
within. Here, $\NPDA$ is the class of pushdown automata, and $\NPCM$ is the class of pushdown automata augmented by some number
of reversal-bounded counters \cite{Ibarra1978}.
\begin{proposition}
There are languages that are inherently ${\cal A}$-ambiguous, but not
inherently ${\cal B}$-ambiguous, for the following pairs ${\cal A}$ and ${\cal B}$.
\begin{itemize}
\item ${\cal A}$ is the class of nondeterministic reversal-bounded checking stack automata (or reversal-bounded stack automata), ${\cal B}$ is the class
of nondeterministic (or deterministic) checking stack (or stack) automata.
\item ${\cal A} = \NCM(1)$ or ${\cal A} = \NCM(1,1)$ and ${\cal B} = \NCM(2,1)$ or ${\cal B} = \DCM(2,1)$.
\item ${\cal A} = \NPDA$ and ${\cal B} = \NPCM$.
\end{itemize}
\end{proposition}
\begin{proof}
For the first point, it suffices to indicate how a deterministic checking stack machine could unambiguously accept $S = R$ (where $R$ is from
Proposition \ref{inherentlyambetol}). Here, on input $a^n b (a^{m_1}b) \cdots (a^{m_k}b) a^n v, n\geq 1, k \geq 0, m_i <n \mbox{~for~} 1 \leq i \leq k, v\in \{a,b\}^*$, $M$ pushes $a^n$ on the stack, and then for each $a^{m_i}$, it verifies
that $m_i < n$, then it verifies that there is $a^n$ on the input, and then it reads the rest of the input and accepts.

For the second point, we use the following well-known inherently $\CFG$-ambiguous language (also inherently ambiguous over nondeterministic pushdown automata): 
$L = \{a^i b^j c^k \mid  i, j \ge 1,  i = j$ or $j = k\}$. $L$ can be accepted by a $\DCM(2,1)$ (and hence
by an unambiguous $\NCM(2,1)$) that reads $a^i$ and stores $i$ to one counter, then reads $a^j$
while decrementing the counter to check if $i = j$, and at
the same time storing $j$ in another counter to check 
if $j = k$ when it reads $a^k$. The third point follows since $L$ is inherently $\NPDA$-ambiguous, but can be
accepted by a $\DCM(2,1)$ (and hence an unambiguous $\NPCM$).
\qed
\end{proof}
To note here, for the latter result involving $\NCM$, even though the language $L$ is in $\NCM(1,1) \subsetneq \NCM(1)$,
the result for $\NCM(1,1)$ is not necessarily strictly stronger than the result for $\NCM(1,1)$ since there are 
machines in $\NCM(1) - \NCM(1,1)$ that could possibly accept $L$ unambiguously.

This result is obviously also true when ${\cal A}$ is either reversal-bounded Turing machines or queue automata, and
${\cal B}$ is either Turing machines or queue automata (which have the power of Turing machines).

Despite there existing inherently $\NCM(1)$-ambiguous languages, we do not yet have a proof that there are inherently $\NCM$-ambiguous languages, and this problem remains open. It should be noted that the approach using counting regularity above will not work for $\NCM$ with two or more counters since even $\DCM(2,1)$ (which are all unambiguous) contain non-counting regular languages \cite{iba-2-journ}. However, we conjecture that the language $S$ above would be such an example. An even more interesting example would be the following language: $L = \{ \#a^{m_1} \# a^{n_1} \# \cdots \# a^{m_k} \# a^{n_k} \mid k \ge 1,
$ each $ m_i, n_i \ge 1, m_i \ne n_i$ for some $i\}$. This can certainly be accepted by a deterministic one counter machine (no reversal bound) by copying each $m_i$ onto the counter, and comparing it to $n_i$, accepting if at lest one is different.
Clearly $L$ can be accepted by an $\NCM(1,1)$ machine.
Although we are currently unable to prove that there is no unambiguous $\NCM$ accepting $L$, we do know that there is no $\DCM$ machine accepting $L$, which can be seen as follows:
\begin{claim}
$L = \{ \#a^{m_1} \# a^{n_1} \# \cdots \# a^{m_k} \# a^{n_k} \mid k \ge 1,
$ each $ m_i, n_i \ge 1, m_i \ne n_i$ for some $i\} \notin \DCM$.
\end{claim}
\begin{proof}
Assume $L$ can be accepted by a a $\DCM$. Let $A$ be the complement of
$L$. It is known that $\DCM$ is closed under complementation and intersection \cite{Ibarra1978}.

Let $B = A \cap R$, where $R$ is the regular set $\{ \#a^{m_1} \# a^{n_1} \# \cdots \# a^{m_k} \# a^{n_k} \mid k \ge 1, \mbox{~each~} m_i, n_i \ge 1\}$.
Then $B = \{\#a^{m_1} \# a^{n_1} \# \cdots \# a^{m_k} \# a^{n_k} \mid k \ge 1, m_i = n_i \ge 1 \mbox{~for each~} i\}$ can be accepted by a $\DCM$ $M_B$.

From $M_B$, we can construct an $\NCM$ $M_C$ accepting 
$C = \{ \# a^s \# a^{m_1} \# a^{n_1} \# \cdots \# a^{m_k} \# a^{n_k} \# a^r \mid k \ge 1, s,r \ge 1, m_i = n_i, \ge 1 \mbox{~for each~} i\}$. On input $a^s \# a^{m_1} \# a^{n_1} \# \cdots \# a^{m_k} \# a^{n_k} \# a^r,  k \ge 1, s,r \ge 1, m_i = n_i, \ge 1$ for each $i$, $M_C$ reads past the first segment $a^s$ and then simulates $M_B$. At some point, when $M_C$ is
scanning $\#$, it guesses that the next $a$-segment is the last one and continues simulating $M_B$ while on $\#$ as if it were the end-marker. If $M_B$
accepts, then $M_C$ verifies that the next $a$-segment (i.e., $a^r$) is indeed the last one and accepts.

So $B$ has an even number of segments, as does $C$. Further, $D = B \cap C = \{ \#(a^n \#)^k \mid n \ge 1, k \mbox{~even} \}$. Furthermore it is known that $\NCM$ is closed under intersection, and
 the Parikh image of any $\NCM$ language is semilinear \cite{Ibarra1978}. However, the Parikh image of $D$
is not semilinear, a contradiction.
\qed
\end{proof}

The authors are also unsure of whether there exist any languages that are inherently ambiguous with respect to the finite-index context-free grammars 
that are not inherently $\CFG$-ambiguous. If the answer is `yes', then what is an example of such a language?

Next, we look at inherent ambiguity of bounded languages. We know that all bounded Ginsburg semilinear languages are in $\DCM$.
Thus, every bounded Ginsburg semilinear language (or every bounded language in any semilinear trio, of which $\NPCM$ is an 
example \cite{Ibarra1978}) can be accepted by an unambiguous $\NCM$ machine. Hence, we have:
\begin{proposition} There are no bounded languages that are inherently $\NCM$-ambiguous (resp.\ inherently $\NPCM$-ambiguous).
\end{proposition}
Is there also a class of grammars where this is true? We show next that this is true:
\begin{proposition} The following are true:
\begin{itemize}
\item All bounded Ginsburg semilinear languages (all bounded languages in any semilinear trio) can be generated by an unambiguous reduced $\ETOL\fin$ system and an unambiguous $\M\fin$ grammar.
\item There are no bounded languages that are inherently $\M\fin$-ambiguous languages or inherently $\ETOL\fin$-ambiguous.
\end{itemize}
\end{proposition}
\begin{proof}
Since the bounded Ginsburg semilinear languages, the bounded languages in any semilinear trio, and the bounded languages generated by both $\ETOL\fin$ and $\M\fin$ coincide (Propositions \ref{thm5}, \ref{equiv-fin-amb-ET0L-MatrGram}, and \ref{ifandonlyif}), it is enough to show that all bounded Ginsburg semilinear languages can be generated by an unambiguous reduced $\ETOL\fin$ system.

Let $L \subseteq w_1^* \cdots w_k^*$ be a bounded Ginsburg semilinear induced by $Q$. By Proposition \ref{thm5}, $L$
is a $\DCM$ language. From Lemma 12 of \cite{iba-2-journ}, there exists a semilinear set $Q$ such that $\phi(Q) = L$, and
$\phi$ is injective on $Q$ (where $\phi$ is the function from Definition \ref{boundedDefs}). Thus, for every
$w \in L$, there is a unique $(l_1, \ldots, l_k) \in Q$ such that $\phi(l_1, \ldots, l_k) = w$.

We know semilinear sets are a finite union of linear sets. A linear set is called {\em simple} if the periods form a basis. A semilinear set is semi-simple if it is the finite disjoint union of simple sets. 
%
%
In view of the Eilenberg-Sch\"utzenberger Theorem, it  is known that given any semilinear set, there is a procedure to effectively construct another set of constants and periods that forms a semi-simple set generating the same set \cite{ES-1969} (see also \cite{Sakarovitch,FlavioSparse}).
Thus, let $Q$ be the disjoint finite union of simple sets $Q_1, \ldots, Q_q$, and let $\vec{v_{i0}}$ be the constant vector of $Q_i$, and let $\vec{v_{ij}}$ be the $j$'th period in the $i$'th set, for $1 \leq j \leq r_i$.
Thus, given any word $w$ in $L$, there is a unique $(l_1, \ldots, l_k) \in Q$ such that $\phi(l_1, \ldots, l_k)=w$, a unique linear set $Q_i$ with $(l_1, \ldots, l_k) \in Q_i$, and $(l_1, \ldots, l_k)$ has a unique combination of the constant and periods of $Q_i$.

Hence, we construct a reduced unambiguous $\ETOL\fin$ system $G$ as follows. First, from the initial nonterminal, $G$
selects immediately guesses a linear set, $Q_i$, from which the yield $w$ will have $\phi^{-1}(w) \in Q_i$ (and there is only one such $Q_i$).  Next, $G$ starts $k$ parallel branches. The $s$'th branch will only generate copies of the word $w_s$, for $1 \leq s \leq k$. Every branch $s$ starts by generating
$\vec{v_{i0}}(s)$ copies of $w_s$.
Then, for each period $\vec{v_{ij}}$, one at a time for $ 1 \leq j \leq r_i$, $G$ guesses some number $x_j$ and every
branch $s$ in parallel generates $x_j \cdot \vec{v_{ij}}(s)$ copies of $w_i$.
Let $(l_1, \ldots, l_k) = \vec{v_{i0}} + x_1 \cdot \vec{v_{i1}} + \cdots + x_{r_i} \cdot \vec{v_{ir_i}}$.
Thus, the yield of such a derivation is
$$w_1^{\vec{v_{i0}}(1)+ x_1 \cdot \vec{v_{i1}(1)} + \cdots + x_{r_i} \cdot\vec{v_{ir_i}}(1)  } \cdots w_k^{\vec{v_{i0}}(k) + x_1 \cdot \vec{v_{i1}}(k) + \cdots + x_{r_i} \cdot\vec{v_{ir_i}}(k)}
 =  w_1^{l_1} \cdots w_k^{l_k}.$$
Thus, $L(G) = \phi(Q) = L$, and each word in $L$ only has one derivation since there is only one combination of the constant and periods giving $(l_1, \ldots, l_k)$.
\qed
\end{proof}

\section{The Commutative Equivalence Problem for   Finite-Index Matrix Languages}\label{sec:Commutative-Equivalence-Problem}
   We consider now the Commutative Equivalence Problem ({\em CE Problem}, for short)   for finite-index matrix languages. 
The notion of commutative equivalence plays an important role in the study of several problems of theoretical computer science such as, for instance, in the theory of codes, where it is involved in the celebrated Sch\"utzenberger conjecture about the commutative equivalence of a maximal finite code with a prefix one (see e.g, \cite{C3.3}).

   The CE Problem investigates the conditions that assure that a language  
   is commutatively equivalent to a regular language. 
       It is worth noticing that commutatively equivalent languages share the same alphabet and their
   generating series are equal. In particular, the characteristic series in commutative variables, and thus the
   generating series, of a commutatively regular language are rational.      This fact implies  
    the answer to the CE Problem is not trivial in general  for finite-index matrix languages due to the existence of languages such as $S$ above whose generating series are algebraic, and not rational. 

Next, we show some conditions that provide a positive answer to the CE Problem 
  for the class of unambiguous finite-index matrix languages.
     For this purpose, some notions and results are needed.

Let $\Sigma$ be an  alphabet.  A subset $W$ of $\Sigma^+$ is a \emph{code (over $\Sigma$)} if every word of $W^+$ has a unique factorization as a product of words of $W$.
 A set $W$ over the alphabet $\Sigma$ is said to be a \emph{prefix code} if $W\Sigma^+\cap W= \emptyset$, that is, if, for every $u, v\in W$, $u$ is not a proper prefix of $v$.

We start by proving the following lemma.
\begin{lemma}
	\label{lemma:lista}
	Let $\mathcal M=(v_1, \ldots, v_m)$ be a list of words in $\Sigma^+$ such that: 
	\begin{enumerate}
		\item for $i=1,\ldots, m$, $|v_i|\geq m$;
		\item for every $a \in\Sigma$, there exists at most one word $v_i\in a^+$. 
	\end{enumerate} 
	Then there exists a prefix code $W=\{w_1,\ldots,w_m\}$ such that $w_i\sim v_i$, $i=1,\ldots,m$.
\end{lemma}

 \begin{proof}
	We proceed by induction on $m$.
	
	If $m=1$, the statement is trivially true.
	Thus, we assume $m\geq 2$.
	With no loss of generality, we suppose that $|v_m|=\max_{1\leq i\leq m}|v_i|$.
	By the inductive hypothesis, 
	there exists a prefix code $\mathcal Y=\{y_1,\ldots,y_{m-1}\}$ such that $y_i\sim v_i$, $i=1,\ldots,m-1$. 
	To prove the statement, it is sufficient to find a word $y_m$ such that $y_m\sim v_m$ and no word of $\mathcal Y$ is a prefix of $y_m$.
	
	Suppose that $v_m=a^n$ for some $a\in \Sigma$, $n\geq m$.
	By Condition (2), no other word of $\mathcal M$, and, consequently, no word of $\mathcal Y$ is a power of $a$.
	Thus, it is sufficient to take $y_m=v_m$.
	
	Now, let us consider the case that $v_m$ is not the power of a single letter.
	Then, we can find a factor $u$ of $v_m$ of length  $m$ containing at least two distinct letters.
	The number of the words which are commutatively equivalent to $u$ is not smaller than $m$.
	Thus, among these words, at least one is different from all the prefixes of length $m$ of the words of $\mathcal Y$.
	Let $v$ be such a word.
	One has $v_m\sim us\sim vs$ for some $s\in \Sigma^*$ and no word of $\mathcal Y$ can be a prefix of $vs$, since otherwise, $v$ would be a prefix of such a word.
	Thus, our goal is attained taking $y_m=vs$.\qed
\end{proof}

\begin{proposition}
	\label{thm-cor1}
	Let $G=(N,\Sigma,M,S)$ be an unambiguous matrix grammar of index $k$ in normal form.
	Assume that there exist a code $W$ and a bijection $f\colon M\to W$ such that, for every matrix $m=(X_1\rightarrow\gamma_1,\ldots,X_h\rightarrow\gamma_h)$ one has $f(m)\sim\pi_{\Sigma}(\gamma_1\cdots \gamma_h)$.
	Then $L(G)$ is commutatively regular. 
\end{proposition}

\begin{proof}
	Since $W$ is a code then the map $f$ can be extended to a monomorphism (an injective morphism) $f\colon M^*\to \Sigma^*$.
	As one easily verifies, for all $\alpha\in M^*$, one has $f(\alpha)\sim\theta(\alpha)$, where $\theta$ is the morphism considered in Lemma \ref{lem:phi}.
	We introduce the map $F\colon L(G)\to W^*$ defined as follows.
	For all  $v\in L(G)$ we set $F(v)=f(\alpha)$, where $\alpha$ is the unique element of $D(G)$ such that $S\derivestep[\alpha]v$.
	
	Taking into account Lemma~\ref{lem:phi}, for all  $v\in L(G)$ one has
	\[
	F(v)=f(\alpha)\sim\theta(\alpha)\sim v,
	\]
	where $\alpha$ is chosen as above.
	Moreover, from Condition P1 and the injectivity of $f$, one has that $F$ is injective.
	We conclude that the set $L(G)$ is commutatively equivalent to $F(L(G))$.
	
	Now, from the definition of $F$, one easily obtains $F(L(G))=f(D(G))$.
	Taking into account that $D(G)$ is a regular set by Proposition~\ref{prop:Proposition15}  and that morphisms preserve regularity, we conclude that $F(L(G))$ is a regular set.
	The statement follows.\qed
\end{proof}

Putting the previous two results together, we obtain:
\begin{proposition}
	\label{thm-cor2}
	Let $G=(N,\Sigma,M,S)$ be an unambiguous finite-index matrix grammar in normal form.
    Suppose that the following conditions are verified: 
	\begin{enumerate}
		\item for every matrix $m=(X_1\rightarrow\gamma_1,\ldots,X_h\rightarrow\gamma_h)$, one has $|\gamma_1\cdots \gamma_h|_\Sigma\geq |M|$.
		
		\item For all letter $a\in\Sigma$, there exists at most one  matrix $m=(X_1\rightarrow\gamma_1,\ldots,X_h\rightarrow\gamma_h)$ such that $\pi_\Sigma(\gamma_1\cdots \gamma_h)\in a^*$. 
	\end{enumerate}
	Then $L(G)$ is commutatively regular. 
\end{proposition}
\begin{proof}
	Clearly, it is sufficient to verify that the hypotheses of Proposition~\ref{thm-cor1} are satisfied.
	This is, in fact, a straightforward consequence of Lemma~\ref{lemma:lista}. 
	\qed
\end{proof}
It is known that a Greibach normal form holds for matrix grammars \cite{PaunMatrix3}, but this only requires that each matrix have at least one production with at least one terminal letter. This is not strong enough for the conditions of this proposition to hold.

We now provide a suitable adaptation of Propositions \ref{thm-cor1} and \ref{thm-cor2}  to finite-index ET0L systems.
Let $G=(V,\mathcal P,S,\Sigma)$ be an unambiguous reduced finite-index ET0L system.
For all $X\in V$, we denote by $R_X$ the set of the right hand sides of all productions $X\to\alpha$ occurring in the production tables of $G$.
 
\begin{proposition}
\label{prefixcode}
	Let $G=(V,\mathcal P,S,\Sigma)$ be an unambiguous reduced finite-index ET0L system.
	Suppose that for all $X\in V$ there exist a prefix code $Y_X\subseteq\Sigma^*$ and a bijection $f_X\colon R_X\to Y_X$ such that for all  $\alpha\in R_X$ one has
	\[f(\alpha)\sim\pi_\Sigma(\alpha).\]
	Then, $L(G)$ is commutatively regular.
\end{proposition}

\begin{proof}
For the sake of brevity, we limit ourselves to give an outline of the proof, omitting some technical details.

	We will construct a regular grammar $G'=(N,\Sigma,P,S')$ generating a language commutatively equivalent to $L(G)$.
	
	First, we associate with any 1-step derivation $x\derivestep[G] y$, $x,y\in(V\cup\Sigma)^*$, a word of $\Sigma^*$.
	If $x\derivestep[G] y$, then one has
	\[
		x = v_0A_1v_1A_2v_2\cdots A_nv_n,\quad 
		y = v_0\alpha_1v_1\alpha_2v_2\cdots\alpha_nv_n,
	\]
	with $v_j\in\Sigma^*$, $A_i\in V$, $\alpha_i$ in $R_{A_i}$, $1\leq i\leq n$, $0\leq j\leq n$.
	With such a derivation, we associate the word
	\[u=f_{A_1}(\alpha_1)f_{A_2}(\alpha_2)\cdots f_{A_n}(\alpha_n).\]
	One can easily verify that $\pi_\Sigma(y)\sim\pi_\Sigma(x)u$ and that the word $y$ is uniquely determined by the knowledge of $x$ and $u$.
	Moreover, if $u_1$ and $u_2$ are associated with two distinct derivations $x\derivestep[G] y_1$ and $x\derivestep[G] y_2$, then $u_1$ cannot be a prefix of $u_2$.
	
	Now, we construct the grammar $G'$.
	Let $k$ be the index of $G$.
	We take $N=\{Z_x\mid x\in V^{\leq k}\}$ and $S'=Z_S$.
	For all derivation $x\derivestep[G] y$, with $x\in V^{\leq k}$ and $y\in (V\cup\Sigma)^*$, $G'$ has the production
	$Z_x\to uZ_{\pi_V(y)},$ where $u$ is the word associated with the derivation $x\derivestep[G] y$, as explained above.
	Also, $Z_\lambda \to\lambda$ is a production.
	
	The regular grammar $G'$ is unambiguous.
	This result can be proved by exploiting the property that
	distinct 1-step derivations of $G$ with the same left side are associated with words where one is not a prefix of the other.
	
	In order to verify that $G$ and $G'$ generate commutatively equivalent languages,
	we associate with any derivation
	\begin{equation}\label{eq:generation}
		S=\alpha_0\derivestep[G]\alpha_1\derivestep[G]\alpha_2\derivestep[G]
		\cdots\derivestep[G]\alpha_n=w,
	\end{equation}
	$w\in L(G)$, the derivation
	\begin{align*}
		Z_S=Z_{\pi_V(\alpha_0)}&\derivestep[G']
		u_1Z_{\pi_V(\alpha_1)}\derivestep[G']
		u_1u_2Z_{\pi_V(\alpha_2)}\derivestep[G']\cdots \derivestep[G']u_1u_2\cdots u_nZ_{\pi_V(\alpha_n)}\derivestep[G']
		u_1u_2\cdots u_n,
	\end{align*}
	where $u_i$ is the word associated to the $i$th step of (\ref{eq:generation}), $1\leq i\leq n$.
	One can verify that, in this way, we have established a 1-to-1 correspondence between the derivations of the words of $L(G)$ and of $L(G')$, such that corresponding derivations produce commutatively equivalent words.
	
	Taking into account that both $G$ and $G'$ are unambiguous, we conclude that they generate commutatively equivalent languages.
	This concludes the proof.$\quad\Box$
\end{proof}

\begin{remark}
The grammar $G'$ of the previous proof generates a regular prefix code.
	Thus, the languages generated by ET0L systems satisfying the hypotheses of Proposition \ref{prefixcode} are commutatively equivalent to regular prefix codes.
\end{remark}

From the previous proposition and Lemma \ref{lemma:lista} one easily derives the following:
\begin{proposition}
	Let $G=(V,\mathcal P,S,\Sigma)$ be an unambiguous reduced finite-index $\ETOL$ system.
	Suppose that the following conditions are verified:
	\begin{enumerate}
		\item for every production $X\to\alpha$ one has $|\alpha|_\Sigma\geq |R_X|$,
		\item for all $a\in\Sigma$ and all $X\in V$, there is at most one production $X\to a^n$, with $n\geq 0$.
	\end{enumerate}
	Then, $L(G)$ is commutatively regular.
\end{proposition}

We finally show that some arguments underlying the proof of the previous result can be utilized to show the 
 commutative regularity of  the languages generated by deterministic $\EOL$ systems.
 
For the next lemma, we use $\EDOL$ rather than reduced $\EDOL$. This is because in a reduced $\EDOL$ system generating a non-empty language, only a single derivation ending in a single word in $L(G)$ would be possible.

\begin{lemma}
	\label{lm:finiteED0L}
	If an $\EDOL$  language contains two  words that are commutatively equivalent, then it is finite.
	Moreover, any ambiguous $\EDOL$  system generates a finite language.	
\end{lemma}

\begin{proof}
	Let $G$ be an $\EDOL$  system and
	\[
		S=\gamma_0\Rightarrow \gamma_1\Rightarrow\cdots
		\Rightarrow \gamma_n\Rightarrow\cdots
	\]
	its unique unending derivation starting by the initial symbol $S$.
	
	If $L(G)$ contains two words that are  commutatively equivalent, then one has $\gamma_i\sim\gamma_j$ for some $i>j>0$.
	One easily derives that $\gamma_{i+1}\sim\gamma_{j+1}$ and, more generally, $\gamma_{i+k}\sim\gamma_{j+k}$ for all $k\geq 0$.
	Thus,  the sequence of the commutation classes of the words $\gamma_n$ is ultimately periodic.
	One derives that $L(G)$ is included in the union of finitely many commutation classes and therefore it is a finite set.
	
	If $G$ is ambiguous, then one has $\gamma_i=\gamma_j$ for some $i\geq j\geq 0$.
	Consequently, by the previous argument, $L(G)$ is finite.\qed
\end{proof}

\begin{proposition}\label{1table-commut-equiv}
	All finite-index $\EDOL$  languages are commutatively regular.
\end{proposition}
\begin{proof}
	Let $L$ be a finite-index $\EDOL$ language.
	With no loss of generality, we assume that $L$ is infinite.
	By Lemma \ref{lm:finiteED0L}, $L$ is generated by an unambiguous finite-index $\EDOL$ system, which by
	Proposition \ref{makereduced} can be generated by an unambiguous reduced finite-index $\EDTOL$ (not $\EDOL$ as
	Proposition \ref{makereduced} introduces another table), which can be generated by an unambiguous finite-index
	matrix grammar in normal form by Proposition \ref{equiv-fin-amb-ET0L-MatrGram} and
	Lemma~\ref{FIML-useful-gr}.

	We introduce the map $F\colon L(G)\to\theta(D(G))$ defined as follows.
	For all  $u\in L(G)$ we set $F(u)=\theta(\alpha)$, where $\alpha$ is the unique element of $D(G)$ such that $S\derivestep[\alpha]u$.
	
	Let us verify that $F$ is injective.
	If one has $F(u)=F(v)$ for some $u,v\in L$, then one has $S\derivestep[\alpha]u$, $S\derivestep[\beta]v$, $\theta(\alpha)=\theta(\beta)$ for some $\alpha,\beta \in D(G)$.
	From Lemma~\ref{lem:phi}, 
	one derives $u\sim v\sim\theta(\alpha)$ and therefore, as $L$ is infinite, from Lemma~\ref{lm:finiteED0L} one obtains $u=v$.
	This proves that $F$ is injective.
	
	Now, the proof can be achieved similarly to that of Proposition~\ref{thm-cor1}. 
	\qed
\end{proof}

%

Despite the commutative regularity of all finite-index $\EDOL$ languages, and the counting regularity of $\LL(\ETOL\fin)$ and $\LL(\M\fin)$, it
is still open whether all languages in $\LL(\ETOL\fin)$ and $\LL(\M\fin)$ are commutatively regular.

\section{Conclusions and Future Directions}
\label{sec:conclusions}

In this paper, it was shown that all bounded languages in any semilinear trio have positive decidability properties, are all
commutatively regular, and are in $\LL(\DCM)$. In particular, the bounded languages in $\LL(\NCM)$ and those generated by finite-index $\ETOL$ systems (and finite-index matrix grammars) coincide. For non-bounded languages, it was shown that all languages generated by
unambiguous finite-index matrix grammars have rational characteristic and generating series, and are counting regular. This implies that there are inherently ambiguous finite-index matrix (and finite-index $\ETOL$) languages. Lastly, the commutative equivalence problem was studied for finite-index matrix and $\ETOL$ languages. In particular, it was shown that all finite-index $\EDOL$ languages are commutatively regular.

Many problems remain open, and there are some interesting future directions. 
First, it is open as to whether there is an $\LL(\NCM)$ language that cannot be generated by a finite-index $\ETOL$ system.
It is also unknown whether there exist any inherently ambiguous $\ETOL$ languages. Furthermore, it is open whether all languages
generated by unambiguous finite-index matrix grammars are commutatively regular.

More generally, it is easily verified that two commutatively equivalent languages, have the same  characteristic series in commutative variables. 
Hence the   characteristic series  of a commutatively regular language, is rational.  
The latter is a necessary condition for commutative regularity, but, as far as we know, it is not sufficient. In this theoretical context,
one may consider the following two questions:
\begin{enumerate}
\item Does there exists a language with a rational characteristic series, whose complement does not have a rational characteristic series?
\item Does there exists a language $L$ such that both $L$ and its complement have rational characteristic series, which is not commutatively regular?
   \end{enumerate}

Since commutatively regular languages are closed under complement, a positive answer to question 1 would show that rationality of the characteristic series is not a sufficient condition for commutative regularity.
A negative answer to question 2 could be viewed as a multidimensional 
version of a recent result of B\'eal and Perrin characterizing generating series of regular languages \cite{BP}.

\bibliography{FiniteIndexBounded}{}

\begin{thebibliography}{10}
\expandafter\ifx\csname url\endcsname\relax
  \def\url#1{\texttt{#1}}\fi
\expandafter\ifx\csname urlprefix\endcsname\relax\def\urlprefix{URL }\fi
\expandafter\ifx\csname href\endcsname\relax
  \def\href#1#2{#2} \def\path#1{#1}\fi

\bibitem{GinsburgCFLs}
S.~Ginsburg, The Mathematical Theory of Context-Free Languages, McGraw-Hill,
  Inc., New York, NY, USA, 1966.

\bibitem{harrison1978}
M.~Harrison, Introduction to Formal Language Theory, Addison-Wesley series in
  computer science, Addison-Wesley Pub. Co., 1978.

\bibitem{Parikh}
R.~Parikh, On context-free languages, J. ACM 13~(4) (1966) 570--581.

\bibitem{IbarraSeki}
O.~H. Ibarra, S.~Seki, Characterizations of bounded semilinear languages by
  one-way and two-way deterministic machines, International Journal of
  Foundations of Computer Science 23~(6) (2012) 1291--1306.

\bibitem{Ibarra1978}
O.~H. Ibarra, Reversal-bounded multicounter machines and their decision
  problems, J. ACM 25~(1) (1978) 116--133.

\bibitem{RozenbergFiniteIndexETOL}
G.~Rozenberg, D.~Vermeir, On {ET0L} systems of finite index, Information and
  Control 38 (1978) 103--133.

\bibitem{PaunMatrix}
G.~P{\u a}un, On the family of finite index matrix languages, Journal of
  Computer and System Sciences 18 (1979) 267--280.

\bibitem{RozenbergFiniteIndexGrammars}
G.~Rozenberg, D.~Vermeir, On the effect of the finite index restriction on
  several families of grammars, Information and Control 39 (1978) 284--302.

\bibitem{Duske}
J.~Duske, R.~Parchmann, Linear indexed languages, Theoretical Computer Science
  32~(1--2) (1984) 47--60.

\bibitem{LATA2017}
F.~D'Alessandro, O.~H. Ibarra, I.~McQuillan, On finite-index indexed grammars
  and their restrictions, in: F.~Drewes, C.~Mart\'in-Vide, B.~Truthe (Eds.),
  Lecture Notes in Computer Science, Vol. 10168 of 11th International
  Conference on Language and Automata Theory and Applications, LATA 2017,
  Ume\r{a}, Sweden, Proceedings, 2017, pp. 287--298.

\bibitem{multipushdown}
L.~Breveglieri, A.~Cherubini, C.~Citrini, S.~Reghizzi, Multi-push-down
  languages and grammars, International Journal of Foundations of Computer
  Science 7~(3) (1996) 253--291.

\bibitem{Harju2002278}
T.~Harju, O.~Ibarra, J.~Karhumaki, A.~Salomaa, Some decision problems
  concerning semilinearity and commutation, Journal of Computer and System
  Sciences 65~(2) (2002) 278--294.

\bibitem{iba}
O.~Ibarra, B.~Ravikumar, On sparseness, ambiguity and other decision problems
  for acceptors and transducers, in: B.~Monien, G.~Vidal-Naquet (Eds.), 3rd
  Annual Symposium on Theoretical Aspects of Computer Science, Vol. 210 of
  Lecture Notes in Computer Science, Springer Berlin Heidelberg, 1986, pp.
  171--179.

\bibitem{BP}
M.-P. B\'eal, D.~Perrin, On the generating sequences of regular languages on
  $k$ symbols, J. ACM 50 (2003) 955--980.

\bibitem{iba-2-journ}
O.~Ibarra, I.~McQuillan, B.~Ravikumar, On counting functions and slenderness of
  languages, Theoret.\ Comput.\ Sci. 777 (2019) 356--378.

\bibitem{Chomsky}
N.~Chomsky, Handbook of Mathematical Psychology, Wiley, New York, 1963, Ch.
  Formal Properties of Grammars.

\bibitem{Blattner}
M.~Blattner, Inherent ambiguities in families of grammars, in: H.~A. Maurer
  (Ed.), Lecture Notes in Computer Science, Vol.~71 of International Conference
  on Automata, Languages, and Programming (ICALP), Springer Berlin Heidelberg,
  1979, pp. 38--48.

\bibitem{GrammarForms}
A.~Cremers, S.~Ginsburg, Context-free grammar forms, Journal of Computer and
  System Sciences 11 (1975) 86--117.

\bibitem{FiniteTurn}
S.~Ginsburg, E.~H. Spanier, Finite-turn pushdown automata, SIAM Journal on
  Control 4~(3) (1966) 429--453.

\bibitem{di2013-code}
F.~D'Alessandro, B.~Intrigila, On the commutative equivalence of bounded
  context-free and regular languages: the code case, Theoret.\ Comput.\ Sci.
  562 (2015) 304--319.

\bibitem{di2013-diof}
F.~D'Alessandro, B.~Intrigila, On the commutative equivalence of semi-linear
  sets of $\natnum ^k$, Theoret.\ Comput.\ Sci. 562 (2015) 476--495.

\bibitem{di2013-sem}
F.~D'Alessandro, B.~Intrigila, On the commutative equivalence of bounded
  context-free and regular languages: the semi-linear case, Theoret.\ Comput.\
  Sci. 572 (2015) 1--24.

\bibitem{CD-DLT-2018}
A.~Carpi, F.~D'Alessandro, On the commutative equivalence of context-free
  languages, in: M.~Hoshi, S.~Seki (Eds.), Developments in Language Theory
  2018, Vol. 11088 of Lecture Notes in Computer Science, Springer Berlin
  Heidelberg, 2018, pp. 429--440.

\bibitem{DassowPaun}
J.~Dassow, G.~P{\u a}un, Regulated Rewriting in Formal Language Theory, Vol.~18
  of EATCS Monographs in Theoretical Computer Science, Springer, 1989.

\bibitem{AmbiguityE0L}
A.~Ehrenfeucht, G.~Rozenberg, On ambiguity in {E0L} systems, Theoretical
  Computer Science 12~(2) (1980) 127--134.

\bibitem{G75}
S.~Ginsburg, Algebraic and Automata-Theoretic Properties of Formal Languages,
  North-Holland Publishing Company, Amsterdam, 1975.

\bibitem{eDCM}
O.~Ibarra, I.~McQuillan, The effect of end-markers on counter machines and
  commutativity, Theoretical Computer Science 627 (2016) 71--81.

\bibitem{RS}
G.~Rozenberg, A.~Salomaa, The Mathematical Theory of L Systems, Academic Press,
  Inc., New York, 1980.

\bibitem{InstructionsTCS}
O.~H. Ibarra, I.~McQuillan, On families of full trios containing counter
  machine languages, Theoretical Computer Science 799 (2019) 71--93.

\bibitem{Baker1974}
B.~S. Baker, R.~V. Book, Reversal-bounded multipushdown machines, Journal of
  Computer and System Sciences 8~(3) (1974) 315--332.

\bibitem{Rozoy}
B.~Rozoy, The {Dyck} language {$D_1^{\prime *}$} is not generated by any matrix
  grammar of finite index, Information and Computation 74~(1) (1987) 64--89.

\bibitem{G78}
S.~Greibach, Remarks on blind and partially blind one-way multicounter
  machines, Theoretical Computer Science 7 (1978) 311--324.

\bibitem{beau}
P.~Beauquier, Deux families de langages incomparables, Inform. and Control 43-2
  (1979) 101--121.

\bibitem{PaunMatrix2}
G.~P{\u a}un, Some further remarks on the family of finite index matrix
  languages, RAIRO --- Informatique Th\'{e}orique 13~(3) (1979) 289--297.

\bibitem{SalomaaSoittola}
A.~Salomaa, M.~Soittola, Automata-theoretic Aspects of Formal Power Series,
  Springer, Berlin, 1978.

\bibitem{Eilenberg}
S.~Eilenberg, Automata, Languages, and Machines (Volume {A}), Vol. 59-A,
  Academic Press, New York, 1974.

\bibitem{BaronKuich}
G.~Baron, W.~Kuich, The characterization of nonexpansive grammars by rational
  power series, Information and Control 48~(2) (1981) 109--118.

\bibitem{InherentAmbiguityE0L}
A.~Ehrenfeucht, G.~Rozenberg, R.~Verraedt, On inherently ambiguous {E0L}
  languages, Theoretical Computer Science 28 (1984) 197--214.

\bibitem{Gross}
M.~Gross, Inherent ambiguity of minimal linear grammars, Information and
  Control 7 (1964) 366--368.

\bibitem{fla}
P.~Flajolet, Analytic models and ambiguity of context-free languages,
  Theoretical Computer Science 49 (1987) 283--309.

\bibitem{CheckingStack}
S.~Greibach, Checking automata and one-way stack languages, Journal of Computer
  and System Sciences 3~(2) (1969) 196--217.

\bibitem{ES-1969}
S.~Eilenberg, M.~P. Sch\"utzenberger, Rational sets in commutative monoids, J.
  of Algebra 13 (1969) 173--191.

\bibitem{Sakarovitch}
J.~Sakarovitch, Elements of Automata Theory, Cambridge University Press, New
  York, NY, USA, 2009.

\bibitem{FlavioSparse}
F.~D'Alessandro, B.~Intrigila, S.~Varricchio, On the structure of the counting
  function of sparse context-free languages, Theoretical Computer Science
  356~(1) (2006) 104--117.

\bibitem{C3.3}
J.~Berstel, D.~Perrin, C.~Reutenauer, Codes and Automata, Encyclopedia of
  Mathematics and its Applications No.~129, Cambridge University Press,
  Cambridge, 2009.

\bibitem{PaunMatrix3}
G.~P{\u a}un, Some context-free-like properties of finite index matrix
  languages, Bulletin Math\'{e}matique de la Soci\'{e}t\'{e} des Sciences
  Math\'{e}matiques de la R\'{e}publique Socialiste de Roumainie 27~(75) (1983)
  83--87.

\end{thebibliography}
\bibliographystyle{elsarticle-num}

\end{document}